\documentclass[a4paper,fleqn,usenatbib]{mnras}

\usepackage[T1]{fontenc}
\usepackage{ae,aecompl}

\usepackage{savesym}
\usepackage{amsmath}
\savesymbol{iint}
\usepackage{txfonts}
\restoresymbol{TXF}{iint}

\usepackage{graphicx}	
\usepackage{amssymb}	

\usepackage{threeparttable} 
\usepackage{threeparttablex} 

\newcommand{\vsi}{\ensuremath{v_{\mathrm{Si}}}}
\newcommand{\vsidot}{\ensuremath{\dot{v}_{\mathrm{Si}}}}
\newcommand{\kms}{\,km\,s$^{-1}$}
\newcommand{\kmsmpc}{\,km\,s$^{-1}$\,Mpc$^{-1}$}
\newcommand{\dmB}{\ensuremath{\Delta m_{15}(B)}}

\title[Early Observations of SN\,2015F]{Early observations of the nearby type Ia supernova SN\,2015F}
\author[Cartier et al.]{R. Cartier$^{1}$\thanks{E-mail: r.cartier-ugarte@soton.ac.uk (RC)}, 
M. Sullivan$^{1}$, R. Firth$^{1}$, G. Pignata$^{2, 3}$, P. Mazzali$^{4, 5}$, K. Maguire$^{6}$,
\newauthor M.~J. Childress$^{1}$, I. Arcavi$^{7, 8}$, C. Ashall$^{4}$, B. Bassett$^{9, 10, 11}$, S.~M. Crawford$^{9}$, C. Frohmaier$^{1}$,
\newauthor  L. Galbany$^{12, 13}$, A. Gal-Yam$^{14}$, G. Hosseinzadeh$^{7, 8}$, D.~A. Howell$^{7, 8}$, C. Inserra$^{6}$, J. Johansson$^{14}$, 
\newauthor E. K. Kasai$^{9,10,11,15}$, C. McCully$^{7, 8}$, S. Prajs$^{1}$, S. Prentice$^{4}$, S. Schulze$^{3,16}$, S.~J. Smartt$^{6}$,
\newauthor K.~W. Smith$^{6}$, M. Smith$^{1}$, S. Valenti$^{7, 8}$, and D.~R. Young$^{6}$ \\
$^{1}$Department of Physics and Astronomy, University of Southampton, Southampton, Hampshire, SO17 1BJ, UK\\
$^{2}$Departamento de Ciencias Fisicas, Universidad Andres Bello, Avda. Republica 252, Santiago, Chile\\
$^{3}$Millennium Institute of Astrophysics, Santiago, Chile\\
$^{4}$Astrophysics Research Institute, Liverpool John Moores University, IC2, Liverpool Science Park, 146 Brownlow Hill, Liverpool L3 5RF, UK\\
$^{5}$Max-Planck-Institut f\"ur Astrophysik, Karl-Schwarzschild-Str. 1, D-85748 Garching, Germany\\
$^{6}$Astrophysics Research Centre, School of Mathematics and Physics, Queens University Belfast, Belfast BT7 1NN, UK\\
$^{7}$Las Cumbres Observatory Global Telescope Network, 6740 Cortona Dr., Suite 102 Goleta, Ca 93117\\
$^{8}$Department of Physics, University of California, Santa Barbara, CA 93106-9530, USA\\
$^{9}$South African Astronomical Observatory, P.O.Box 9, Observatory 7935, South Africa\\
$^{10}$African Institute for Mathematical Sciences, 6-8 Melrose Road, Muizenberg 7945, South Africa\\
$^{11}$Department of Mathematics and Applied Mathematics, University of Cape Town, Rondebosch, 7700, South Africa\\
$^{12}$Pittsburgh Particle Physics, Astrophysics, and Cosmology Center (PITT PACC).\\ 
$^{13}$Physics and Astronomy Department, University of Pittsburgh, Pittsburgh, PA 15260, USA.\\
$^{14}$Department of Particle Physics and Astrophysics, Weizmann Institute of Science, Rehovot 76100, Israel\\
$^{15}$Department of Physics, University of Namibia, 340 Mandume Ndemufayo Avenue, Pioneerspark, Windhoek, Namibia\\
$^{16}$Instituto de Astrof\'isica, Facultad de F\'isica, Pontificia Universidad Cat\'olica de Chile, Vicu\~na Mackena 4860, 7820436 Macul, Santiago, Chile\\
}
\date{Accepted 2016 October 14. Received 2016 October 11; in original form 2016 May 25}

\pubyear{2016}

\begin{document}
\label{firstpage}
\pagerange{\pageref{firstpage}--\pageref{lastpage}}
\maketitle

\begin{abstract}
  We present photometry and time-series spectroscopy of the nearby
  type Ia supernova (SN Ia) SN\,2015F over $-16$ days to $+80$ days
  relative to maximum light, obtained as part of the Public ESO
  Spectroscopic Survey of Transient Objects (PESSTO).  SN\,2015F is a
  slightly sub-luminous SN Ia with a decline rate of
  $\dmB=1.35\pm0.03$\,mag, placing it in the region between normal and
  SN\,1991bg-like events.  Our densely-sampled photometric data place
  tight constraints on the epoch of first light and form of the
  early-time light curve. The spectra exhibit photospheric \ion{C}{ii}
  $\lambda 6580$ absorption until $-4$\,days, and high-velocity
  \ion{Ca}{ii} is particularly strong at $<-10$\,days at expansion
  velocities of $\simeq$23000\kms.  At early times, our spectral
  modelling with \textsc{syn++} shows strong evidence for iron-peak
  elements (\ion{Fe}{ii}, \ion{Cr}{ii}, \ion{Ti}{ii}, and \ion{V}{ii})
  expanding at velocities $>14000$\kms, suggesting mixing in the
  outermost layers of the SN ejecta.  Although unusual in SN Ia
  spectra, including \ion{V}{ii} in the modelling significantly
  improves the spectral fits.  Intriguingly, we detect an absorption
  feature at $\sim$6800\,\AA\ that persists until maximum light. Our
  favoured explanation for this line is photospheric \ion{Al}{ii},
  which has never been claimed before in SNe Ia, although detached
  high-velocity \ion{C}{ii} material could also be responsible. In
  both cases the absorbing material seems to be confined to a
  relatively narrow region in velocity space.  The nucleosynthesis of
  detectable amounts of \ion{Al}{ii} would argue against a
  low-metallicity white dwarf progenitor. We also show that this
  6800\,\AA\ feature is weakly present in other normal SN Ia events,
  and common in the SN\,1991bg-like sub-class.
\end{abstract}


\section{Introduction}
\label{sec:introduction}

The uniformity of type Ia supernova (SN Ia) light curves allows them
to be used as reliable distance indicators, providing crucial evidence
for the accelerated expansion of the universe \citep{riess98,
  perlmutter99}. Despite many years of research and the general
agreement that the progenitor stars of SNe Ia are accreting
carbon-oxygen (CO) white dwarfs in binary systems, the nature of the
companion star \citep{2014ARA&A..52..107M}, and the detailed physics
of the explosion, remain uncertain.

The study of the outer layers of SN Ia ejecta can, in principle,
provide important clues about the progenitor white dwarf and explosion
physics by tracing the extent and amount of any unburnt material and
the metallicity of the progenitor star \citep{hoflich98, lentz00,
  walker12, maguire12, foley13, mazzali14}.  In particular, early
ultraviolet (UV) spectra are sensitive to the abundance of iron-group
elements in the outermost layers, and can place important constraints
on progenitor metallicity \citep{hachinger13, maguire12, foley13,
  mazzali14}.
Any carbon detected in the outermost layers is particularly important,
as carbon is the only element that could not be the result of
thermonuclear burning, and can be directly associated with the
original composition of the CO white dwarf.
The amount and distribution of carbon can place strong constraints on
the extension of the burning front and the degree of mixing during the
explosion \citep{branch03, thomas07, parrent12}.

These outer layers can only be studied with early spectroscopic
observations. The unburned material can be detected as absorption
lines of \ion{C}{ii} in the optical \citep{parrent11, thomas11b,
  folatelli12, silverman12, maguire14, cartier14}, and of \ion{C}{i} in the
near-infrared \citep[NIR; ][]{hoflich02, marion06, marion09, hsiao13, hsiao15, marion15}.
Recent studies have shown that at least 30 per cent of
SNe Ia possess \ion{C}{II} absorption lines prior to maximum light
\citep{thomas11b, folatelli12, silverman12, maguire14}.

Early spectra of SNe Ia also commonly exhibit `high-velocity' (HV)
features. These spectroscopic features correspond to absorption lines
with expansion velocities much higher than the photospheric velocity,
and usually greater than 15000\kms, sometimes reaching 30000\kms\ or
higher at the earliest phases.  The most common HV features are of
\ion{Ca}{II}, which seem to be a ubiquitous phenomenon at early stages
\citep{mazzali05, childress14, maguire14, silverman15}. HV features of
\ion{Si}{II} are rarer \citep[see][]{marion13, childress13,
  silverman15}, and HV features of other ions (\ion{S}{II},
\ion{Fe}{II}, \ion{C}{II}, \ion{O}{I}) have also been claimed
\citep{fisher97, hatano99, mazzali01, branch03, garavini04, nugent11,
  marion13, cartier14}.

Such high expansion velocities suggest that HV features are produced
in the outermost layers of the SN ejecta.  Therefore, it is reasonable
to hypothesize that their origin is tightly linked to the progenitor
system and/or the physics of the burning in the outermost layers of
the white dwarf.  HV features are ubiquitous in SN Ia spectra at about
a week prior to maximum light \citep{mazzali05, marion13, childress14,
  maguire14, silverman15, zhao15}, and decrease in strength with time
\citep{maguire14, silverman15, zhao15}.  Possible explanations for HV features
include density enhancements from swept-up \citep{gerardy04} or
distant \citep{tanaka06} circumstellar material, abundance
enhancements in the outermost layers of the ejecta \citep{mazzali05},
or variations of the ionization state in the outer layers due to non
`local thermodynamic equilibrium' (LTE) effects \citep{blondin13}.
Their origin remains a puzzle.

The advent of high-cadence wide-area sky surveys over the last ten
years has meant that the quality and quantity of early SN discoveries
has increased, and with it has come a wealth of early SN Ia
spectroscopy.  In this paper, we present spectroscopy and photometry
of the nearby SN Ia SN\,2015F. In Section~\ref{observations_sec}, we
introduce SN\,2015F and describe the photometry and spectroscopy,
beginning at $-16.30$\,d relative to peak brightness and extending to
$+75.5$\,d past peak. We also estimate the distance to the host of
SN\,2015F (NGC 2442), the rise time, and the epoch of first light.
In Section~\ref{analysis_sec}, we analyse the
spectroscopic data, and in Section \ref{spec_mod_sec} we model the
spectra using the {\sc syn++} code. We discuss our results in
Section~\ref{discusion_sec}, and summarize in
Section~\ref{conclusions_sec}. Throughout, we assume a value for the
Hubble constant of $H_0=70$\,km\,s$^{-1}$\,Mpc$^{-1}$.

\section{Observations}
\label{observations_sec}

\begin{figure*}
\centering
\includegraphics[width=170mm]{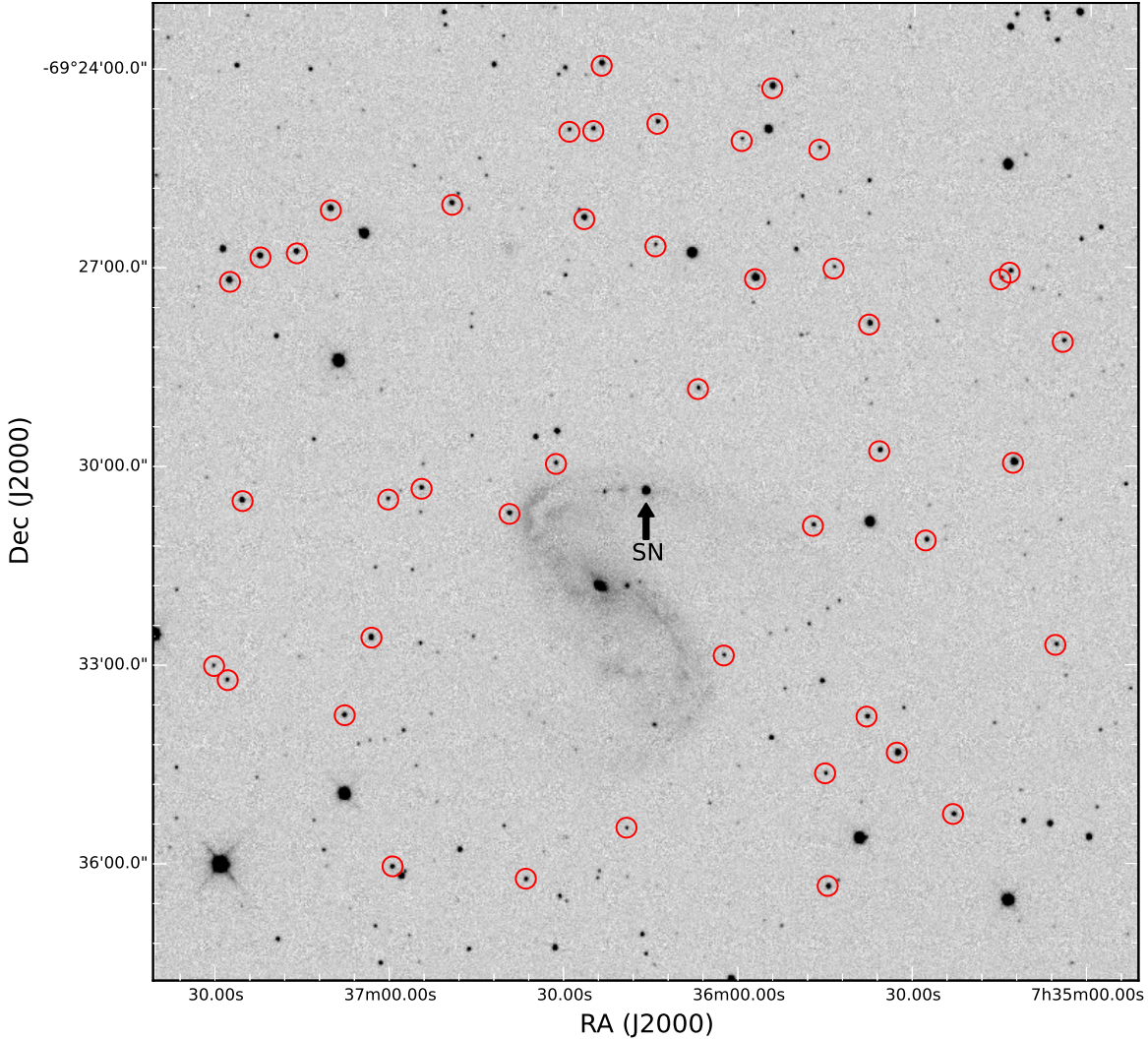}
\caption{$V$-band image of the field of SN\,2015F obtained with the
  LCOGT 1-m telescope at the Siding Spring Observatory, Australia.
  The SN position is highlighted with an arrow. The stars of the photometric
  sequence around the SN, which were used to obtain differential
  photometry of the SN, are indicated with a red circle.}
\label{SN2015F_chart_fig}
\end{figure*}

SN\,2015F is located $43\farcs4$ north and $86\farcs2$ west of the
centre of the nearby spiral galaxy NGC~2442, on the northern arm of NGC~2442,
and was discovered on 2015 March 09.789 (all dates are UT) by
\citet{monard15} at $\alpha = 07^{\mathrm{h}}36^{\mathrm{m}}15\fs76$,
$\delta = -69\degr30\arcmin23\farcs0$ (see Fig.~\ref{SN2015F_chart_fig}). 
The unfiltered discovery magnitude was 16.8. SN\,2015F was promptly
classified as a young SN Ia by the Public ESO Spectroscopic Survey
of Transient Objects (PESSTO) collaboration on March 11.00
\citep{fraser15}. NGC~2442 is a SBbc galaxy with a recession velocity of
$1466\pm5$\kms\ in HyperLeda\footnote{\url{http://leda.univ-lyon1.fr/}}, and the Milky
Way reddening along the line-of-sight to SN\,2015F is
$E(B-V)_{\mathrm{MW}}=0.175$\,mag \citep{schlafly11}, corresponding to
a $V$-band extinction ($A_V$) of $\simeq$0.54\,mag. 

The classification of SN\,2015F as a young SN sparked a detailed
spectroscopic and photometric follow-up campaign, which we detail in
this section.

\subsection{Photometry}
\label{sec:photometry}

We used several instruments to obtain photometry of SN\,2015F, the
main characteristics of which are summarized in
Table~\ref{cameras_tab}.  Data obtained with the European Southern
Observatory (ESO) Faint Object Spectrograph and Camera (v2) (EFOSC2)
on the New Technology Telescope (NTT) were reduced with the PESSTO pipeline
described in detail in \citet{smartt15}, and the Las Cumbres Observatory Global
Telescope Network (LCOGT) images were reduced using the Obsevatory
Reduction and Acquisition Control Data Reduction pipeline
\citep[ORAC-DR][]{jenness15}. The reduction steps of the data obtained
with the PROMPT telescopes are described in \citet{pignata11}.

\begin{table*}
  \caption{Summary of imagers used to observe SN\,2015F.}
\label{cameras_tab}
\begin{tabular}{@{}lcccc}
\hline
Site & Telescope/instrument & Detector Size & Plate scale           & Filters \\
     &                      &               & (arcsec pixel$^{-1}$) & \\
\hline
CTIO     & PROMPT/CCD          & $1024 \times 1024$ & $0.600$ & $BV$ \\
La Silla & NTT/EFOSC           & $1024 \times 1024$ & $0.240$ & $UBVgr$ \\
CTIO     & 1-m~LCOGT--005/SBIG & $2048 \times 2048$ & $0.467$ & $UBVgri$ \\ 
SSO      & 1-m~LCOGT--003/SBIG & $2048 \times 2048$ & $0.467$ & $UBVgri$ \\ 
SAAO     & 1-m~LCOGT--010/SBIG & $2048 \times 2048$ & $0.467$ & $UBVgri$ \\ 
SSO      & 1-m~LCOGT--011/SBIG & $2048 \times 2048$ & $0.467$ & $UBVgri$ \\ 
SAAO     & 1-m~LCOGT--012/SBIG & $2048 \times 2048$ & $0.467$ & $UBVgri$ \\ 
SAAO     & 1-m~LCOGT--013/SBIG & $2048 \times 2048$ & $0.467$ & $UBVgri$ \\ 
La Silla & NTT/SofI            & $1048 \times 1048$ & $0.288$ & $JHK_{s}$ \\
\hline
\end{tabular}
\end{table*}

A optical photometric sequence of stars around the SN was calibrated
in the $UBV$ and $gri$ bandpasses against standard stars from the \citet{landolt92}
and \citet{smith02} catalogues, respectively.
The magnitudes of the optical photometric sequence
around SN\,2015F (see Fig.~\ref{SN2015F_chart_fig}) are presented
in Table~\ref{op_local_seq_tab} in the Appendix.
Point spread function (PSF) fitting photometry was performed on the SN 
frames using \textsc{daophot} \citep{stetson87}, and calibrated using the
photometric sequence around the SN. The optical photometry
of SN\,2015F is in Table~\ref{op_phot_tab}.

The NIR data were obtained with the Son of Isaac (SofI) camera on the
NTT, and were reduced using our own \textsc{IRAF}\footnote{IRAF is
  distributed by the National Optical Astronomy Observatory, which is
  operated by the Association of Universities for Research in
  Astronomy, Inc., under cooperative agreement with the National
  Science Foundation.} scripts to create a clean sky image, which we
subtract from the science images.  Pixel-to-pixel variations were
removed by dividing the science images by a flat field image.  We used
\textsc{scamp} \citep{bertin06} to obtain an astrometric solution, and
\textsc{swarp} \citep{bertin02} to combine the dithered images into a
single image. The NIR photometric sequence was calibrated by observing
\citet{persson98} standard fields close in time and in airmass
to the SN observations. The photometry of the NIR photometric sequence
and of SN\,2015F are presented in Tables \ref{nir_local_seq_tab} and
\ref{nir_phot_tab}, respectively.

\begin{figure*}
\includegraphics[width=180mm]{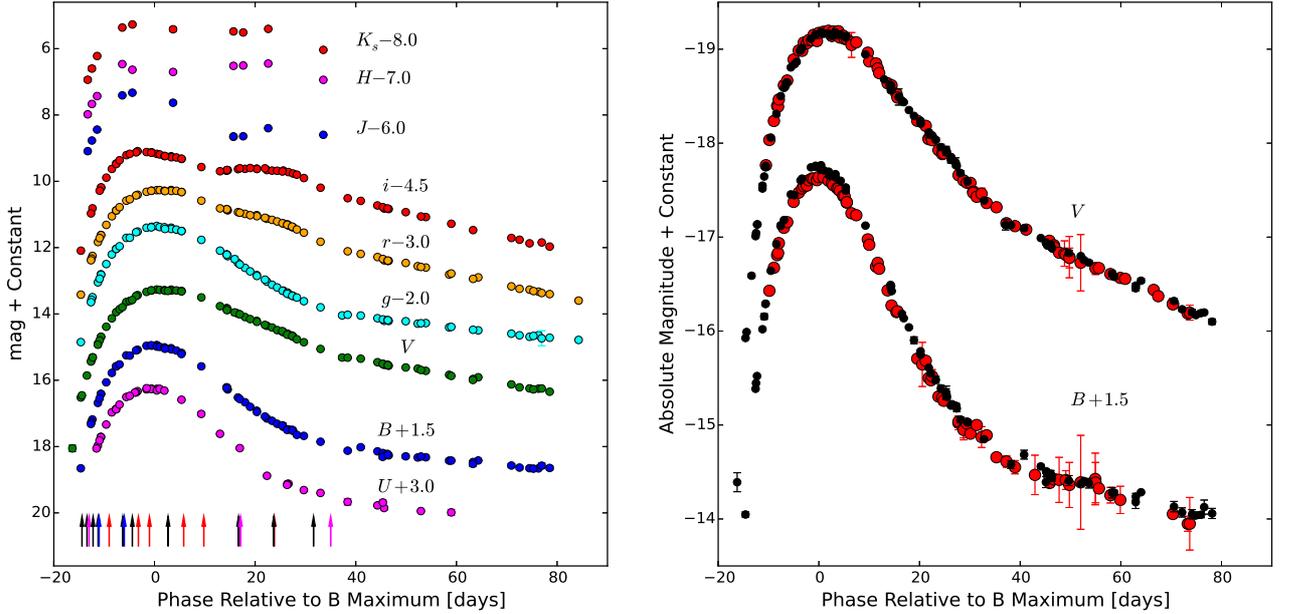}
\caption{{\it Left-panel:} The observed $UBVgriJHK_{s}$ light
  curves of SN\,2015F. The light curves are offset as indicated for
  clarity. The vertical arrows indicate the epochs of the optical
  spectra, and are colour-coded to indicate different instruments as in
  Fig.~\ref{spec_sec_fig}. No corrections for extinction have been made. 
  {\it Right-panel:} A comparison between the $BV$ absolute magnitude light curves
  of SN\,2015F (black circles) and SN\,2004eo \citep[red points;][]{pastorello07}.
  We corrected both SNe for extinction; for SN\,2004eo we assumed $E(B-V)$ values from
  \citet{pastorello07}. Both SNe show remarkably similar light curves, and the main
  difference between them is that SN\,2004eo is slightly fainter by $\sim 0.1$
  mag in $M_{B}$.}
\label{SN2015F_lc_fig}
\end{figure*}

The $UBV$, $gri$ and $JHK_{s}$ light curves of SN\,2015F are shown in
Fig.~\ref{SN2015F_lc_fig}. Fitting a polynomial to the $B$-band light
curve of SN\,2015F, we measure a decline rate of
$\dmB=1.35\pm0.03$\,mag, and we estimate the epoch of maximum light as
MJD $57106.45\pm0.02$\,d (2015 March 25.4). Using the SiFTO light
curve fitter \citep{2008ApJ...681..482C}, we determine a stretch of
$s=0.906 \pm 0.005$.  A summary of the epoch of peak brightness, the
peak magnitude, and the decline rate of the light curve ($\Delta
m_{15}$) obtained from fitting a polynomial to the light curves of
SN\,2015F in each filter is presented in Table~\ref{peak_mag_tab}.

In Table~\ref{reddening_tab} we summarize the values of the host
galaxy reddening along the line-of-sight to SN\,2015F
($E(B-V)_{\mathrm{host}}$), calculated from the optical colours
\citep{phillips99}, and $V-$NIR colours \citep{krisciunas04} of
SN\,2015F. The $E(B-V)_{\mathrm{host}}$ values derived using different
methods are in good agreement with each other, and the weighted
average is $E(B-V)_{\mathrm{host}} = 0.085 \pm 0.019$\,mag; the
uncertainty corresponds to the standard deviation from the weighted
average.

\begin{table}
\caption{Peak magnitude information for SN\,2015F.}
\label{peak_mag_tab}
\begin{tabular}{@{}lccc}
\hline
Filter & MJD peak & Peak & $\Delta m_{15}$ \\
       & (days)   & magnitude  &  \\
\hline
$U$    & $57106.15$($0.06$) & $13.25$($0.03$) & $1.55$($0.04$) \\
$B$    & $57106.45$($0.02$) & $13.46$($0.03$) & $1.35$($0.03$) \\
$V$    & $57108.36$($0.03$) & $13.27$($0.02$) & $0.76$($0.02$) \\
$g$    & $57107.20$($0.01$) & $13.38$($0.03$) & $1.00$($0.03$) \\
$r$    & $57107.88$($0.01$) & $13.26$($0.01$) & $0.68$($0.01$) \\
$i$    & $57104.28$($0.02$) & $13.61$($0.01$) & $0.54$($0.01$) \\
$J$    & $57102.30$($1.00$) & $13.32$($0.06$) & --             \\
$H$    & $57100.23$($1.00$) & $13.46$($0.06$) & --             \\
$K_{s}$& $57103.91$($1.00$) & $13.25$($0.06$) & --             \\
\hline
\end{tabular}
\begin{tablenotes}
\item Numbers in parenthesis correspond to 1-$\sigma$ statistical uncertainties.
\end{tablenotes}
\end{table}

\begin{table*}
  \caption{Host galaxy reddening along the line-of-sight to SN\,2015F.}
\label{reddening_tab}
\begin{tabular}{@{}lcc}
\hline
Method & $E(B-V)_{\mathrm{host}}$ & Reference \\
\hline
$E(B-V)_{\mathrm{tail}}$              & $0.110$($0.050$) & \citet{phillips99} \\
$B_{\mathrm{max}} - V_{\mathrm{max}}$ & $0.066$($0.039$) & \citet{phillips99} \\
$V - H$                               & $0.080$($0.037$) & \citet{krisciunas04} \\
$V - K$                               & $0.093$($0.038$) & \citet{krisciunas04} \\
\hline
Mean                                  & $0.085$($0.019$) & \\
\hline
\end{tabular}
\end{table*}

\subsubsection{Distance to NGC~2442}
\label{sec:distance-ngc-2442}

\begin{table*}
  \caption{Distance estimates to NGC~2442.}
 \label{dist_mod_tab}
 \begin{tabular}{@{}lccc}
  \hline
  Reference & Distance modulus   & Distance & Method/  \\
            & $\mu$ &  (Mpc)  & $\Delta m_{15}$/$M_{\mathrm{max}}$ relation \\
   \hline
   This paper   & $31.63$($0.20$) & $21.2$($2.0$) & SNe Ia/\citet{phillips99} $B-$band calibration \\
   This paper   & $31.64$($0.18$) & $21.3$($1.8$) & SNe Ia/\citet{phillips99} $V-$band calibration \\
   This paper   & $31.68$($0.18$) & $21.7$($1.8$) & SNe Ia/\citet{kattner12} $J-$band calibration \\  
   This paper   & $31.69$($0.14$) & $21.8$($1.4$) & SNe Ia/\citet{kattner12} $H-$band calibration \\
   \citet{im15} & $31.89$($0.04$) & $23.9$($0.4$) & SNe Ia/ $B-$band using MLCS2k2 \citep{jha07} \\
   \citet{tully09} & $31.66$($0.17$) & $21.5$($1.7$) & Tully-Fisher relation \\
   \citet{tully88} & $31.16$($0.80$) & $17.1$($6.3$) & Tully-Fisher relation \\      
  \hline
 \end{tabular}
\end{table*}

We used the peak magnitudes of SN\,2015F to estimate the distance to
its host galaxy, NGC\,2442 (Table~\ref{dist_mod_tab}). We corrected
the observed peak magnitudes for Milky way and host-galaxy extinction
assuming a \citet{cardelli89} reddening law (with $R_{V} = 3.1$). We
estimated the absolute peak magnitudes of SN\,2015F using the
\citet{phillips99} and \citet{kattner12} decline rate/peak luminosity
calibrations in the optical and in the NIR, respectively. We assumed
$H_{0}=70$\kmsmpc\ and, somewhat arbitrarily, an uncertainty of
$3.0$\kmsmpc, placing $H_{0}$ and its uncertainty between the two
currently popular values of $67.3 \pm 1.2$\kmsmpc\ \citep{planck14}
and $73.03 \pm 1.79$\kmsmpc\ \citep{riess16}.  The 1$\sigma$
uncertainty in $H_{0}$ corresponds to an uncertainty of $\simeq4$ per
cent in distance ($0.1$\,mag).

Table~\ref{dist_mod_tab} also lists distance estimates from the
literature, including \citet{im15} distance estimate to
NGC\,2442 also using SN\,2015F. \citet{im15} measured a host
galaxy extinction of $E(B-V)_{\mathrm{host}} = 0.035\pm0.033$\,mag, a
peak magnitude of $B_{\mathrm{max}} = 13.36 \pm 0.10$\,mag, and a
decline rate of $\dmB=1.26 \pm 0.10$\,mag using {\sc snana}
\citep{kessler09} to fit MLCS2k2 templates \citep{jha07} to their
observed data of SN\,2015F. These light curve parameters are
consistent with our values at the 1- to 2-$\sigma$ level. 

Our mean distance modulus in the optical ($BV$-filters) is
$\mu_{\mathrm{optical}} = 31.64\pm0.14$, and in the NIR ($JH$-filters)
is $\mu_{\mathrm{NIR}}=31.68\pm0.11$, where the uncertainties include
filter-to-filter peak magnitude covariances.  These are in excellent
agreement, and are consistent with the \citet{tully09} distance
estimation based on the Tully-Fisher relation. Our mean distance
estimates in the optical and NIR are $1.7$- and $1.8$-$\sigma$
discrepant from the \citet{im15} distance modulus value,
respectively. \citet{im15} quote an uncertainty of $0.04$\,mag
in the distance modulus to NGC\,2442, which appears underestimated
when compared with their quoted uncertainty in the $B$-band peak
magnitude ($0.10$\,mag), and the typical dispersion in the
absolute magnitudes of SNe Ia in the optical \citep[$0.12$ to
$0.16$\,mag;][]{folatelli10}.

\subsubsection{Rise time and Epoch of First Light}
\label{rise_time_sec}

Early observations are fundamental to place constraints
on several properties of SNe Ia, such as the time of the explosion
\citep{nugent11,hachinger13,zheng13,mazzali14,goobar14,zheng14,marion16,shappee16},
the radius of the progenitor \citep{nugent11, bloom12}, and to search for signs
of, or rule out an interaction of the SN ejecta with a companion star
\citep{kasen10,hayden10b,bianco11,brown12,goobar14,cao15,olling15,goobar15,im15,marion16,shappee16}.
Here, using the early $V$-band observations of SN\,2015F, we place strong constraints
on the epoch of first light,  the time when the first photons
diffuse out from the SN ejecta. Recent very early abundance tomography of
SN\,2010jn \citep{hachinger13} and SN\,2011fe \citep{mazzali14} shows
that this estimated epoch of  first light is in tension with the time
of the explosion derived from spectral modelling, implying a dark phase
for these two SNe of the order of $\simeq1$\,d between the time of the explosion
and the emergence of the first photons.

A non-detection of SN\,2015F in the $R$-band was reported
by \citet{im15} on MJD 57088.511, and the first
unambiguous detection (\textgreater3-$\sigma$) on MJD 57089.463,
22.84\,h later. \citet{im15} also discuss a possible 2-$\sigma$
detection of emission from the cooling of the shocked heated ejecta
\citep{piro10, rabinak11} around three days before the first clear detection.
Here, we report 3-$\sigma$ non-detections to a limiting magnitude of $19.017$ and
$18.709$ in the $V$-band on MJD 57089.073 and MJD 57089.184, 9.35\,h
and 6.69\,h before the first detection of \citet{im15} in the
$R$-band. Our first detection of $V = 18.055 \pm 0.101$ is on MJD
57090.124, 15.87\,h after the first detection of \citet{im15}, and
22.56\,h after our last non-detection.

\begin{figure}
\includegraphics[width=85mm]{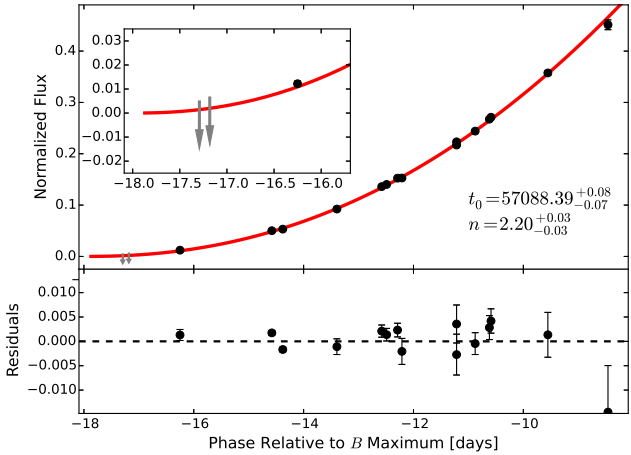}
\caption{In the top-panel, black points denote the normalized early-time
$V$-band light curve  of SN\,2015F; the red line corresponds to the best
fitted power law model \citep[$f_{\mathrm{norm}} \propto (t-t_{0})^{n}$; see][and Section
\ref{rise_time_sec}]{firth15}. Grey arrows correspond to the 3-$\sigma$ upper limits. The 
lower-panel, shows the residuals from the fit. The SN brightness
predicted by the model at the epochs of the upper-limits is consistent with the
non-detections (see inset).}
\label{rise_time_fig}
\end{figure}

We use a parameterization of $f_{\mathrm{model}} = \alpha
(t-t_{0})^{n}$ to fit the rising $V$-band light curve, described in
detail in \citet{firth15}, where $t_{0}$ is the time of first light,
$\alpha$ is a normalizing coefficient, and $n$ is the index of the
power law. The case of $n = 2$ is known as the `expanding fireball'
model \citep[e.g.,][]{riess99}. The early $V$-band light curve,
together with the best fit model, is shown in
Fig.~\ref{rise_time_fig}.

We find best fit parameters of $t_{0} = 57088.394^{+0.076}_{-0.074}$,
and $n = 2.20^{+0.03}_{-0.03}$.  The $V$-band rise time, corrected
for time dilation, is $t_{\mathrm{rise}} = 19.87 \pm 0.08$\,d. The
mean values obtained by \citet{firth15}, for a sample of 18 SNe Ia
with well-sampled early light curves from the Palomar Transient
Factory \citep[PTF][]{law09,rau09} observed in the $R_{\mathrm{P48}}$-band,
and from La Silla-QUEST \citep{baltay13} in the broad $(g+r)$-band
\citep[see][]{baltay13,cartier15}, are $\bar{t}_{\mathrm{rise}} =
18.98 \pm 0.54$\,d, and $\bar{n} = 2.44 \pm 0.13$. \citet{firth15}
showed that the $t_{\mathrm{rise}}$ and $n$ in the $(g+r)$-band is
very similar to an optical pseudo-bolometric optical light curve and
the $R_{\mathrm{P48}}$-band has a $t_{\mathrm{rise}}$ longer by $0.39$\,d.
\citet{firth15} also find that $t_{\mathrm{rise}}$ is longer by 
$0.79$\,d and $n$ is larger by $0.1$ in the $V$-band, compared to an optical
pseudo-bolometric light curve. After applying these corrections, the
$t_{\mathrm{rise}}$ found for SN\,2015F is close to the mean value of
\citet{firth15} sample, and $n$ is $\sim$2\,$\sigma$ lower than the
mean value, although still within the range of values found by
\citet{firth15}.

The early photometry of SN\,2015F does not show any evidence
of shock cooling, such as an excess emission or bluer than
normal colours, produced by the interaction of the SN ejecta
with a companion star as in SN\,2012cg \citep{marion16}. We will
analize in detail possible progenitor scenarios of SN\,2015F using
the very early phase data presented here in a future work.

\subsection{Spectroscopy}
\label{sec:spectroscopy}

We obtained optical spectra of SN\,2015F with EFOSC2/NTT, with 
the Robert Stobie Spectrograph \citep[RSS][]{kobulnicky03} on the
Southern African Large Telescope (SALT), with WiFeS
\citep{dopita09} on the Australian National University (ANU) 2.3-m
Telescope, and with the FLOYDS spectrograph on the 2-m Faulkes
Telescope South (FTS) at the Siding Spring Observatory. The EFOSC2
spectra were reduced as described in \citet{smartt15}, and the FLOYDS
data were reduced with the {\sc floydsspec}
pipeline\footnote{\url{https://www.authorea.com/users/598/articles/6566}}.
The RSS is a long slit spectrograph, the spectra of SN\,2015F were reduced
using the {\sc pysalt} pipeline which is described in detail in \citet{crawford10}.
WiFeS is an integral field unit spectrograph, one dimensional spectra
were reduced and extracted from the data cube using a PSF-weighted fit
using {\sc pywifes} pipeline \citep{2014Ap&SS.349..617C}.  In
Table~\ref{spec_summary_tab} we summarize the optical spectroscopic
observations of SN\,2015F. We correct all our spectra for Galactic
extinction using a \citet{cardelli89} extinction law with $R_{V}=3.1$,
and convert to the rest-frame using the recession velocity of the host
galaxy. All reduced PESSTO data will be available from the ESO Science
Archive Facility in Spectroscopic Survey Data Release 3 (SSDR3), due for
submission in late 2016. Details will be posted on the
PESSTO website\footnote{\url{http://www.pessto.org}}.
All our spectra are available from the WISeREP
archive\footnote{\url{http://wiserep.weizmann.ac.il/}}
\citep{2012PASP..124..668Y}.

\begin{table*}
  \caption{Summary of spectroscopic observations of SN\,2015F.}
 \label{spec_summary_tab}
 \begin{tabular}{@{}lcccccc}
   \hline
   Date UT & MJD        & Phase  & Instrument/ & Wavelength   & Resolution & Exposure \\
   & (days)     & (days) & telescope   & range (\AA)  &  (\AA)     & time (s)      \\
   \hline
   20150311 & $57092.0$ & $-14.4$ & EFOSC/NTT & $3646$ - $9238$ & 18 & 300 \\
   20150312 & $57093.0$ & $-13.4$ & EFOSC/NTT & $3346$ - $9987$ & 14 & 1500 \\   
   20150312 & $57093.4$ & $-13.0$ & WiFeS/ANU 2.3-m & $3500$ - $9565$& 1.2 & 1200 \\
   20150313 & $57094.2$ & $-12.2$ & EFOSC/NTT & $3343$ - $9985$ & 14 & 1500 \\
   20150314 & $57095.2$ & $-11.2$ & EFOSC/NTT & $3343$ - $9983$ & 14 & 600 \\
   20150314 & $57095.4$ & $-11.0$ & FLOYDS/FTS & $3200$ - $9997$ & 12 & 1800 \\
   20150316 & $57097.4$ & $-9.0$ & RSS/SALT & $3875$ - $8091$ & 18 & 330 \\
   20150319 & $57100.1$ & $-6.3$ & EFOSC/NTT & $3343$ - $9985$ & 14 & 600 \\
   20150319 & $57100.4$ & $-6.0$ & FLOYDS/FTS & $3201$ - $9985$ & 12 & 1800 \\
   20150321 & $57102.0$ & $-4.4$ & EFOSC/NTT & $3346$ - $9986$ & 14 & 600 \\
   20150322 & $57103.3$ & $-3.2$ & RSS/SALT & $3875$ - $8091$ & 18 & 300 \\
   20150324 & $57105.4$ & $-1.0$ & RSS/SALT & $3875$ - $8091$ & 18 & 300 \\
   20150328 & $57109.1$ & $+2.7$ & EFOSC/NTT & $3346$ - $9985$ & 14 & 600 \\
   20150331 & $57112.3$ & $+5.8$ & RSS/SALT & $3875$ - $8091$ & 18 & 300 \\
   20150404 & $57116.3$ & $+9.8$ & RSS/SALT & $3875$ - $8091$ & 18 & 300 \\
   20150411 & $57123.1$ & $+16.7$ & EFOSC/NTT & $3343$ - $9983$ & 14 & 900 \\
   20150411 & $57123.5$ & $+17.1$ & WiFeS/ANU 2.3-m & $3500$ - $9565$ & 1.2 & 2400 \\
   20150418 & $57130.1$ & $+23.7$ & EFOSC/NTT & $3343$ - $9984$ & 14 & 900 \\
   20150418 & $57130.2$ & $+23.8$ & RSS/SALT & $3875$ - $8091$ & 18 & 300 \\
   20150426 & $57138.0$ & $+31.6$ & EFOSC/NTT & $3346$ - $9986$ & 14 & 900 \\
   20150429 & $57141.4$ & $+35.0$ & WiFeS/ANU 2.3-m & $3500$ - $9565$ & 1.2 & 2400 \\
  \hline
 \end{tabular}
\end{table*}

\section{Spectroscopic Analysis}
\label{analysis_sec}

\begin{figure*}
\includegraphics{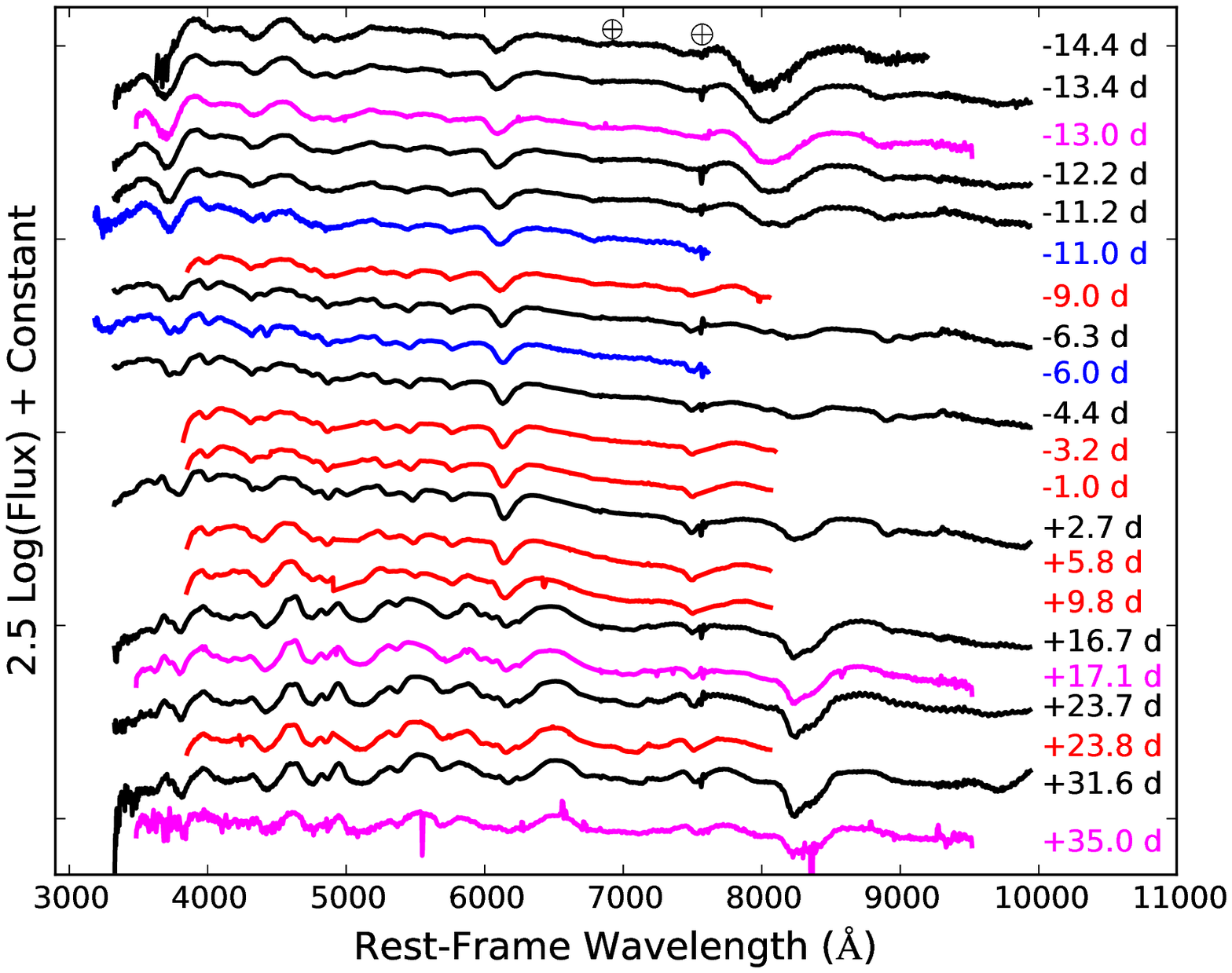}
\caption{The spectral sequence of SN\,2015F. In black we show the
  spectra obtained with EFOSC2/NTT with PESSTO, in red the spectra
  obtained using the RSS spectrograph on SALT,
  in magenta the spectra obtained with WiFeS on the ANU 2.3-m, and in blue the spectra
  obtained using FLOYDS on the Faulkes Telescope South by the LCOGT
  network. The phase of the spectrum relative to maximum light in the
  $B$-band is shown on the right. The position of the main telluric
  features are also marked. In all figures, the spectra have been
  corrected for galactic extinction, aditionally SN\,2015F has been
  also corrected by host galaxy extinction (Section~\ref{sec:photometry}).}
\label{spec_sec_fig}
\end{figure*}

In Fig.~\ref{spec_sec_fig} we show the spectral sequence of SN\,2015F
spanning $-14$ to $+35$\,d (throughout, all phases are given relative
to maximum light in the rest-frame $B$-band). The overall
characteristics of SN\,2015F are those of a normal, if slightly
sub-luminous, SN Ia, with properties particularly similar to
SN\,2004eo (see Fig \ref{SN2015F_lc_fig}), a transitional object between normal and sub-luminous SNe
Ia \citep{pastorello07}. The ratio of the pseudo-equivalent widths
(pEW) of the \ion{Si}{ii} $\lambda 5972$ and \ion{Si}{ii} $\lambda
6355$ features \citep[$\mathcal{R}(\ion{Si}{ii})$; see][]{nugent95,
  bongard06, hachinger08} is $\mathcal{R}(\ion{Si}{ii})=0.31$,
measured from the spectrum obtained at $+2$\,d.  For our $\Delta
m15(B)=1.35$, this is consistent with published relationships between
$\mathcal{R}(\ion{Si}{ii})$ and \dmB\
\citep[e.g.,][]{benetti05,blondin12}.

\subsection{Expansion velocities of Si\,II}
\label{sec:expans-veloc}

\begin{figure}
\includegraphics[width=90mm]{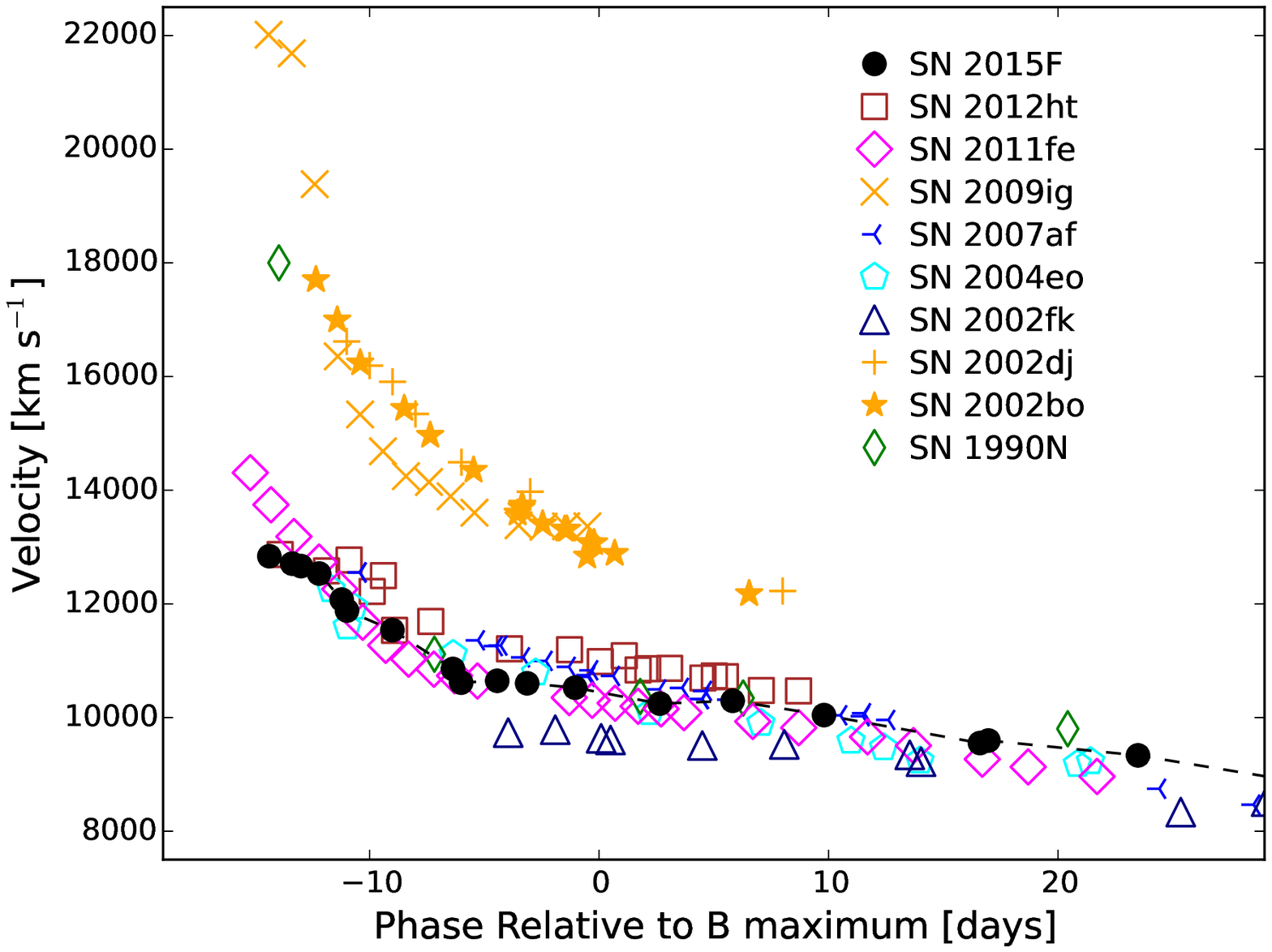}
\caption{\ion{Si}{ii} velocities (\vsi) measured from the minimum of
  the \ion{Si}{ii} $\lambda 6355$ absorption feature for SN\,2015F
  (black circles and dashed line). For comparison we show \vsi\ for
  the LVG SNe Ia SN\,1990N \citep[green diamonds;][]{leibundgut91,
    mazzali93}, SN\,2004eo \citep[cyan pentagons;][]{pastorello07,
    folatelli13}, SN\,2007af \citep[blue;][]{blondin12, folatelli13},
  SN\,2011fe \citep[magenta diamonds;][]{pereira13}, and SN\,2012ht
  \citep[brown squares;][]{yamanaka14}. We also show \vsi\ for the HVG
  SNe Ia SN\,2002bo \citep[orange stars;][]{benetti04}, SN\,2002dj
  \citep[orange plus symbols;][]{pignata08}, and SN\,2009ig
  \citep[orange crosses;][]{foley12, marion13}, and for SN\,2002fk
  \citep[dark blue triangles;][]{cartier14}, characterized by
  persistent \ion{C}{ii} absorption features until past maximum light,
  and by a low \vsi\ and \vsidot.  }
\label{si_ii_comp_velplot_fig}
\end{figure}

In Fig.~\ref{si_ii_comp_velplot_fig} we present the \ion{Si}{ii}
$\lambda 6355$ velocity (\vsi) measured from the minimum of the
$\lambda6355$\,\AA\ absorption feature observed near 6150\,\AA. In the
figure, we compare with several low-velocity gradient (LVG) and
high-velocity gradient (HVG) SNe Ia\footnote{\citet{benetti05}
  separated SNe Ia into three groups: High-velocity gradient (HVG)
  events, consisting of objects with a velocity gradient of
  \ion{Si}{ii} $\lambda 6355$ $\vsidot\geq70$\kms\,d$^{-1}$ and
  $\dmB\leq 1.5$, low-velocity gradient (LVG) events, consisting of
  objects with $\vsidot\leq70$\kms\,d$^{-1}$ and $\dmB\leq1.5$, and
  FAINT events with $\dmB\geq1.5$.}. We estimate \vsi\ at maximum
light ($v^{0}_{\mathrm{Si}}$) and \vsidot, by fitting a first degree
polynomial to the \vsi\ measurements over $-7$ to $+30$\,d, and
interpolating to obtain $v^{0}_{\mathrm{Si}} = 10400$\kms\ and
$\vsidot=50$\kms\,d$^{-1}$. These values place SN\,2015F in the LVG
group.

At phases prior to $-10$\,d, SN\,2015F presents a slower \vsi\ than
SN\,2011fe (see Fig.~\ref{si_ii_comp_velplot_fig}).  The difference in
\vsi\ between these two SNe at $-14$\,d is $\sim$1000\kms.  After
$-10$\,d, SN\,2015F shows a \vsi\ evolution similar to SN\,2011fe and
SN\,2004eo.

\subsection{Comparison to other SNe Ia}
\label{sec:comp-other-supern}

We next compare in detail the spectra of SN\,2015F to other SNe Ia
from the literature. We focus our analysis at phases earlier than
$-10$\,d, and also make a comparison to sub-luminous SN\,1991bg-like
SNe Ia.

\subsubsection{Comparison at $-14$ days}
\label{sec:comparison-at-minus14}

\begin{figure*}
\includegraphics[width=180mm]{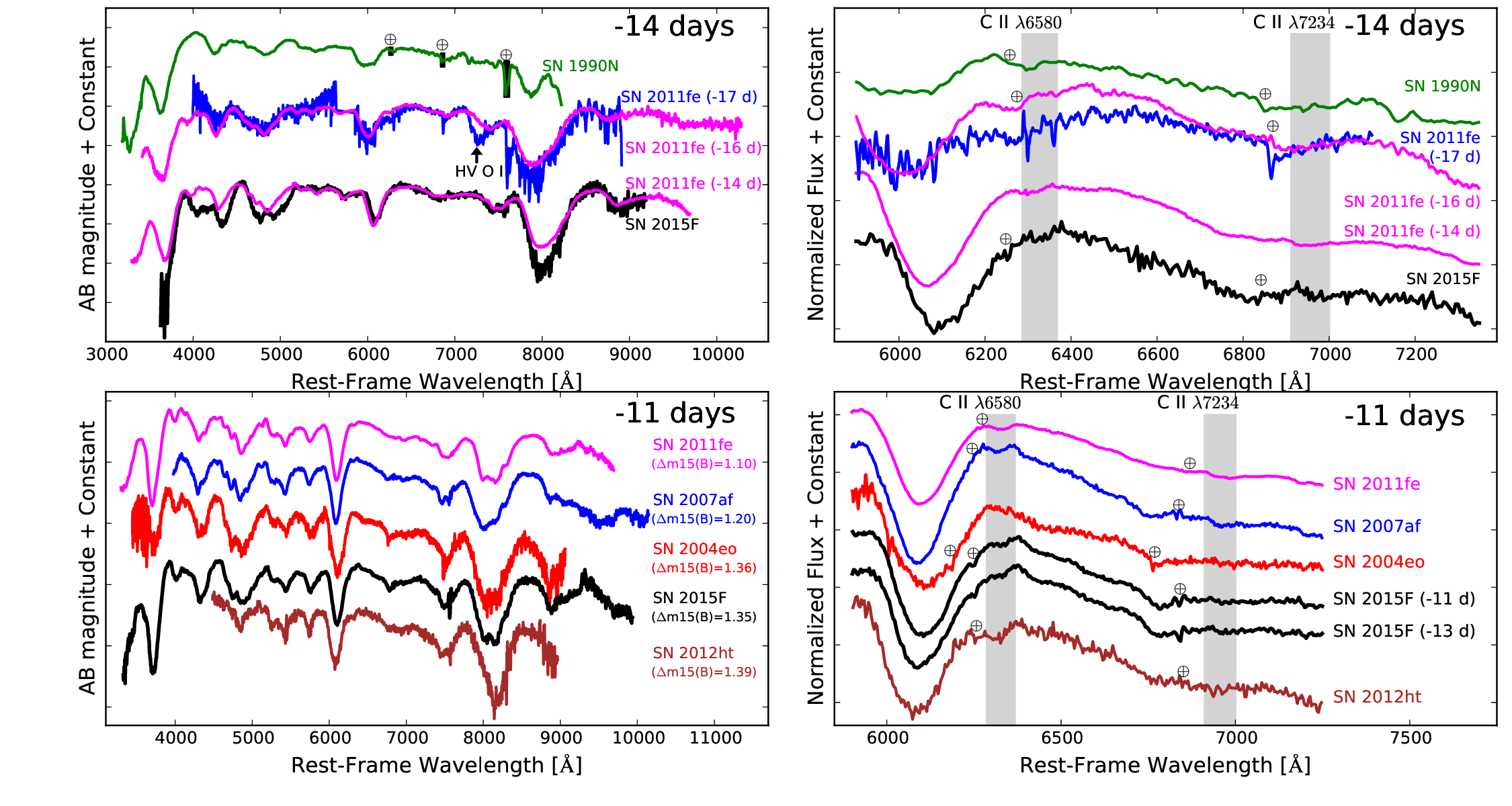}
\caption{{\it Top left:} The spectra of SN\,2015F (black), SN\,1990N
  \citep[green;][]{leibundgut91}, and SN\,2011fe
  \citep[magenta;][]{pereira13} at $-14$\,d. We show for comparison
  the spectra of SN\,2011fe at $-16.6$\,d and $-16.3$\,d in blue and
  magenta \citep{nugent11}, respectively, overplotted to emphasise the
  high-velocity \ion{O}{i} feature, marked with an arrow.  We
  overplotted SN\,2011fe on top of SN\,2015F to highlight the
  similarities and differences. Telluric features are marked with an
  Earth symbol, and in SN\,1990N a black rectangle denotes their
  characteristic width.  {\it Top right:} As top-left, but focussing
  on the region around \ion{C}{ii} $\lambda6580$ and $\lambda 7234$
  at $-14$\,d.  {\it Lower left:} A
  comparison between the spectra of SN\,2015F, SN\,2004eo
  \citep[red;][]{pastorello07}, SN\,2007af
  \citep[blue;][]{folatelli13}, SN\,2011fe
  \citep[magenta;][]{pereira13}, and SN\,2012ht
  \citep[brown;][]{yamanaka14} at $\simeq -11$\,d. The decline rate
  \dmB\ is indicated on the right.  {\it Lower right:} as lower left,
  but focussing on the spectral region around \ion{C}{ii}
  $\lambda6580$ and $\lambda 7234$. In the right panels, the grey 
  region marks the position of photospheric \ion{C}{ii} at
  an expansion velocity of $10000$--$14000$\kms.}
\label{comp_all_fig}
\end{figure*}

The top-left panel of Fig.~\ref{comp_all_fig} compares the first
spectrum of SN\,2015F at $-14$\,d, with SN\,2011fe and SN\,1990N at a
similar phase. We also show the spectra of SN\,2011fe at $-16.6$\,d
and $-16.3$\,d \citep{nugent11}. SN\,2011fe is a normal brightness SN
Ia \citep[$\dmB=1.10$;][]{pereira13} while SN\,1990N is relatively
bright among the group of normal (i.e., non SN\,1991T-like) SNe Ia,
with $\dmB=1.03$ \citep{lira98}. Most of the spectral differences can
be explained as a mere temperature effect. SN\,1990N is brighter/hotter,
and shows fewer features in its spectrum, SN\,2011fe is at an
intermediate luminosity and shows a broadly similar spectrum to
SN\,2015F, while SN\,2015F itself is fainter/cooler and shows more
absorption lines below $5500$\,\AA\ due to singly ionized iron-peak
elements.

The top-right panel of Fig.~\ref{comp_all_fig} shows the same spectra
in the region around the \ion{C}{ii} $\lambda 6580$ and $\lambda 7234$
lines. In SN\,2011fe, a photospheric component of both of these
\ion{C}{ii} lines is clearly observed at $\simeq$13000\kms\
\citep[studied in][]{parrent12}. In SN\,1990N, the lines have also
been reasonably securely identified \citep[e.g.,][]{mazzali01}.
\citet{fisher97} proposed that the flat-bottomed absorption feature at
$\sim$6000\,\AA\ could also be due to \ion{C}{ii} $\lambda 6580$, this
time at high velocity.  Although \citet{mazzali01} showed that
\ion{Si}{ii} is responsible for most of this feature \citep[and in
particular the extended blue side; see also][]{mazzali93}, an
additional explanation may be needed for the red side, which could
include a carbon shell at $20000$\kms.
The SN\,2015F spectrum also shows clear absorption to the red side of
the \ion{Si}{ii} $\lambda 6355$ line, consistent with \ion{C}{ii}
$\lambda 6580$ at $\sim 13000$\kms, a velocity similar to the
photospheric \ion{C}{ii} lines in SN\,2011fe at the same phase
\citep{parrent12}.  The corresponding photospheric \ion{C}{ii}
$\lambda 7234$ can also be seen, although this feature is not strong
in SN\,2015F.

A broad absorption feature at $\sim6800$\,\AA\ is very clear
(top-right Fig.~\ref{comp_all_fig}). Possible identifications include
photospheric \ion{Al}{ii}, or \ion{C}{ii} $\lambda 7234$ at
$20000$\kms.
There is no indication of a corresponding \ion{C}{ii} $\lambda 6580$
at about $20000$\kms, although the inclusion of \ion{C}{ii} in our
modelling yields a slightly better fit to the red side of the
\ion{Si}{II} $\lambda 6355$ absorption line \citep[see
also][]{cartier14}. We model the spectra in detail in
Section~\ref{spec_mod_sec}, where we consider both possibilities.

\subsubsection{Comparison at $-11$ days}
\label{comp_m11_sec}

The lower-left panel of Fig.~\ref{comp_all_fig} shows the comparison
of the LVG SNe Ia SN\,2015F, SN\,2011fe, SN\,2007af, SN\,2004eo and
SN\,2012ht at around $-11$\,d. SN\,2015F shows a remarkable similarity
to SN\,2004eo, with the main difference being the absence of clear
\ion{C}{ii} lines in SN\,2004eo \citep{pastorello07, mazzali08}. A
strong feature in all the spectra is the broad \ion{Ca}{ii} NIR
triplet, extending from $\sim$10000 to $\sim$25000\kms\ (see
Section~\ref{HV_CaII_sec}), with \ion{Ca}{ii} HK also visible where
the spectra extend to the blue. An apparent evolution in the shape and
strength of the spectral features with brightness/temperature (\dmB)
is seen in the region dominated by iron-peak elements (wavelengths
bluer than $\sim5500$\,\AA) and in the \ion{Ca}{ii} NIR triplet;
however, overall there is a remarkable degree of spectral similarity
in this sample of LVG SNe Ia.

In Fig.~\ref{comp_all_fig} we also show the region around \ion{C}{ii}
$\lambda 6580$ and $\lambda 7234$. SN\,2015F, SN\,2012ht, SN\,2011fe
and SN\,2007af show clear photospheric \ion{C}{ii} $\lambda 6580$,
with SN\,2011fe, SN\,2007af, and possibly SN\,2015F showing clear
photospheric \ion{C}{ii} $\lambda 7234$ absorption (the spectrum of
SN\,2012ht is too noisy to clearly detect any weak feature). There is
also clear absorption at $\sim$6800\,\AA\ present in SN\,2015F,
SN\,2007af and perhaps SN\,2004eo, although in this latter case nearby
telluric features hamper a convincing identification.

\subsubsection{Comparison with SN\,1991bg-like type Ia SNe}
\label{sec:comparison-with-sn}

\begin{figure*}

\includegraphics[width=180mm]{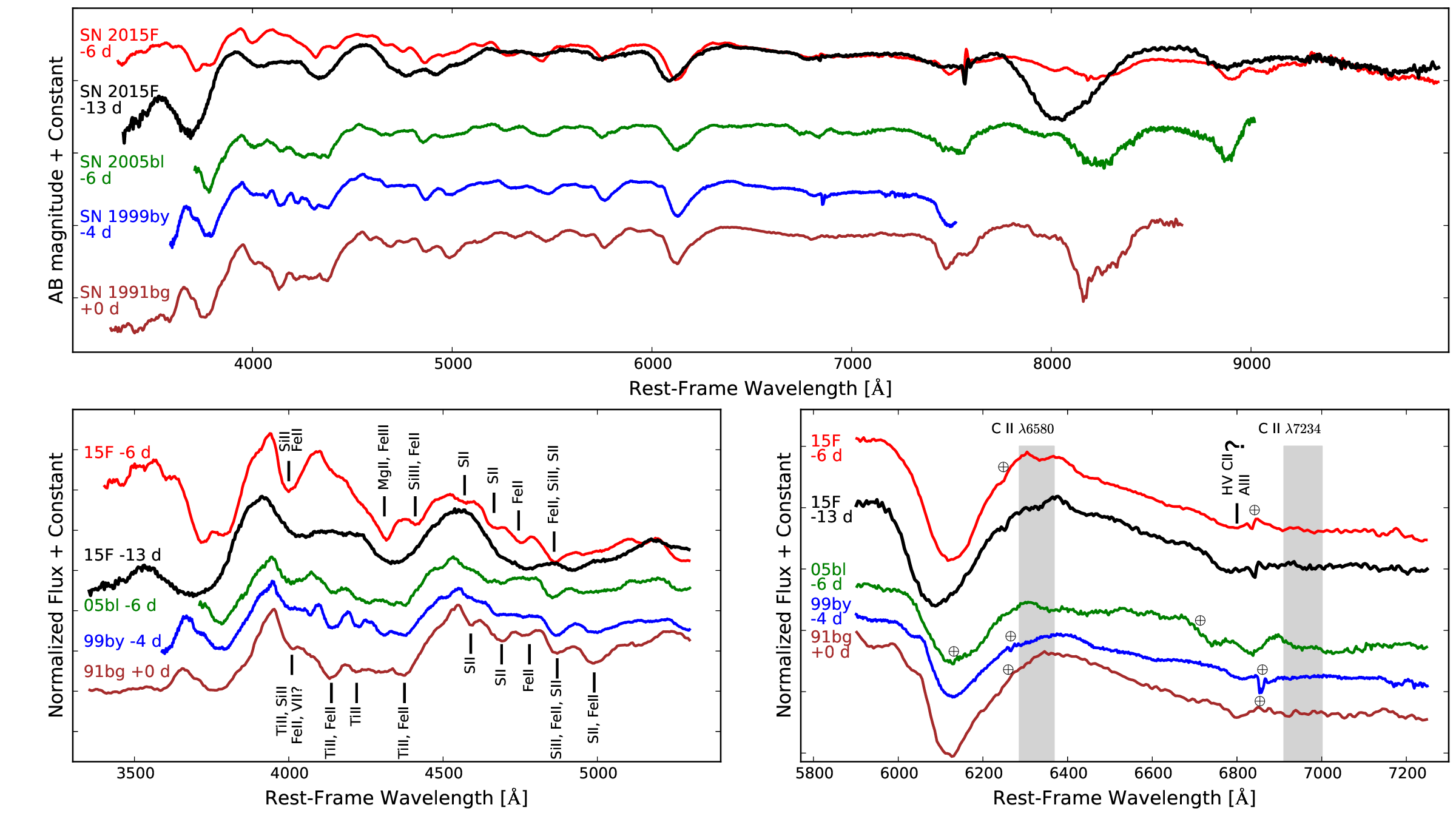}
\caption{{\it Top :} The comparison between SN\,2015F at -13\,d
  (black), SN\,2015F at -6\,d (grey), SN\,1999by at -4\,d
  \citep[blue;][]{garnavich04}, and SN\,1991bg at maximum light
  \citep[brown;][]{filippenko92}. {\it Lower-left:} As in the top
  panel, but now centred on the region 3300--5400\,\AA\ and including
  the SN\,1991bg-like SN\,2005bl at -6\,d
  \citep[green;][]{taubenberger08}. In this spectral region, SNe Ia
  display several lines of iron-peak elements, and \ion{Ti}{ii} lines
  are a distinctive feature of SN\,1991bg-like SNe near to maximum
  light.  {\it Lower right:} As lower left, but focussing on the
  region around the $6800$\,\AA\ feature. The ions responsible for
  prominent spectral features are indicated.}
\label{comp_91bgs_fig}
\end{figure*}

SN\,2015F has some similiarities with SN\,2004eo
(Fig.~\ref{SN2015F_lc_fig} and \ref{comp_all_fig}), a transitional
object between normal and sub-luminous SNe Ia \citep{pastorello07,
  mazzali08}. The SN\,Ia sub-luminous class, often referred to as
SN\,1991bg-like SNe, are characterized by a fast decline in their
light curves \citep{filippenko92}, somewhat lower expansion velocities
compared to normal SNe Ia \citep{hachinger09, doull11}, a small amount
of $^{56}$Ni synthesized during the explosion \citep{mazzali97,
  hoflich02, hachinger09}, and clear \ion{Ti}{ii} lines around maximum
light \citep{filippenko92, mazzali97, garnavich04, taubenberger08,
  doull11}.

Fig.~\ref{comp_91bgs_fig} compared SN\,2015F to a group of
well-observed SN\,1991bg-like events. Although SN\,2015F at -13\,d has
higher expansion velocities than SN\,1991bg-like SNe, with SN\,2015F
showing very broad features formed by the blending of several lines,
it shares some spectral resemblance to SN\,1991bg-like events over
3900--5000\,\AA. This similarity is a consequence of absorption lines
of iron-peak elements such as \ion{Ti}{ii}, \ion{V}{ii} and
\ion{Cr}{ii}, with some of these ions commonly identified in
SN\,1991bg-like SNe \citep[see][]{doull11}.

In Fig. \ref{comp_91bgs_fig} we overplot the spectrum of SN\,2015F at
-6\,d on top of the -13\,d spectrum to highlight the evolution of the
spectral features.  We show in detail the blue part of the spectra,
which are dominated by lines of iron-peak elements. In contrast to the
spectrum obtained at -13\,d that shows \ion{Ti}{ii}, \ion{V}{ii}, and
\ion{Cr}{ii} lines, by -6\,d lines of \ion{Fe}{iii} and \ion{Si}{iii}
are present, commonly seen in normal SNe Ia and which imply a higher
ionization of the SN ejecta compared to previous epochs.

Fig.~\ref{comp_91bgs_fig} also shows a comparison focussed on the
6800\,\AA\ feature. Although sometimes weak, this always appears
present in SN\,1991bg-like objects. The presence of \ion{Ti}{ii} and
\ion{Cr}{ii}, associated with a low ejecta temperature, may also
suggest that the 6800\,\AA\ feature is a product of a low ejecta
temperature.  However, we note that this feature is also present at
-6\,d and -4\,d in SN\,2015F, when there are also clear lines of
doubly-ionized species (\ion{Fe}{iii}, \ion{Si}{iii}), implying a
relatively high ejecta temperature. Thus temperature is unlikely to be
the only parameter that explains the 6800\,\AA\ feature.

The overall appearance of the spectra of SN\,2015F around maximum
light shows that although SN\,2015F shares some similarities with
SN\,1991bg-like SNe, it should be considered as a member of the group
of normal SNe Ia.

\subsection{Ca\,II high-velocity features}
\label{HV_CaII_sec}

\begin{figure}
\includegraphics[width=90mm]{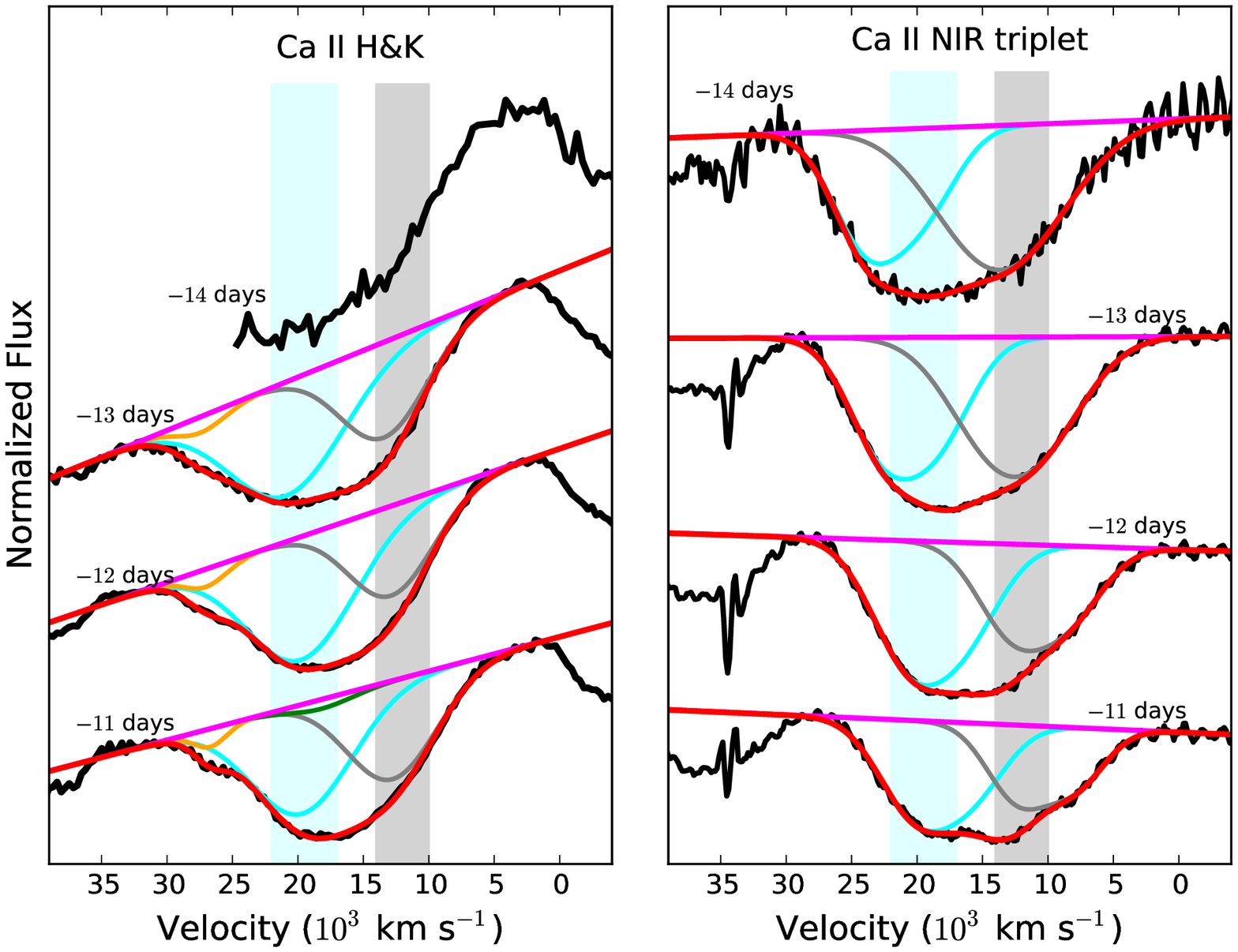}
\caption{The spectral region around the \ion{Ca}{ii} H\&K lines (left
  panel) and the NIR triplet (right panel) for SN\,2015F at phases \textless $-10$\,d.
  The magenta line corresponds to the pseudo-continuum, and the red
  profile is the resultant model from fitting Gaussian profiles to the
  observed spectrum. The profiles of the individual HV and
  photospheric components are shown in cyan and grey, respectively. In
  the left-panel the \ion{Si}{ii} $\lambda 3856$ line is shown in
  green, and the $\sim$3600\,\AA\ feature in orange. The grey and cyan
  regions mark the positions of the photospheric
  ($10000$--$14000$\kms) and HV ($17000$--$22000$\kms) components.}
\label{caii_velplot_fig}
\end{figure}

The \ion{Ca}{ii} high velocity (HV) features are strong in the very
early spectra of SN\,2015F, dominating the \ion{Ca}{ii} features until
a week before maximum light. After these phases the photospheric
component becomes dominant, and the HV component disappears around
maximum light. In Fig.~\ref{caii_velplot_fig} we show the \ion{Ca}{ii}
H\&K and the NIR triplet in velocity space for the $-14$\,d to
$-11$\,d spectra.  For \ion{Ca}{ii} H\&K, the expansion velocity is
calculated with respect to the average wavelength of the H\&K lines
($3951$\,\AA), while for the NIR triplet the velocity is with respect
to the strongest line ($8542$\,\AA).

Following \citet{childress13, childress14}, \citet{maguire14} and
\citet{silverman15}, we fit Gaussian profiles to the \ion{Ca}{ii}
lines, defining a pseudo-continuum on either side of the profile, and
using it to normalize the spectrum. To model the \ion{Ca}{ii} NIR
feature, we used both HV and photospheric components each composed of
three Gaussian profiles with fixed relative positions and a common
width. We allowed the relative strength of the three lines to vary,
but to be the same for both components. The best-fitting parameters
are listed in Table~\ref{caii_NIR_gauss_tab} and are shown in
Fig.~\ref{caii_velplot_fig}. The relative strengths are similar across
the different epochs, and close to the theoretical values expected
from atomic physics.

\begin{table*}
 \caption{Parameters of the Gaussian fits to the \ion{Ca}{ii} NIR triplet.}
 \begin{tabular}{@{}lccccccccc}
   \hline
   Phase   & \multicolumn{3}{c}{HVF Component} & \multicolumn{3}{c}{Photospheric Component} & \multicolumn{2}{c}{Relative Strengths of}  & $R_\mathrm{HVF}$ \\
   &  $v$  &  FWHM  &  pEW   &  $v$  &  FWHM  &  pEW & Ca\,II($\lambda 8542$)/ & Ca\,II($\lambda 8498$)/  & \\
   ~[Days]  &  [km\,s$^{-1}$] & [km\,s$^{-1}$] & [\AA] & [km\,s$^{-1}$] & [km\,s$^{-1}$] & [\AA] & Ca\,II($\lambda 8662$) & Ca\,II($\lambda 8662$)  & \\
   \hline
   $-14.4$ & $23236.0$ & $6425.3$ & $171.6$ & $14794.5$ & $9979.5$ & $238.2$ & $1.31$ & $0.23$ & $0.72$ \\
   $-13.4$ & $22109.1$ & $6548.3$ & $177.1$ & $14049.0$ & $8251.9$ & $197.5$ & $1.09$ & $0.10$ & $0.90$ \\
   $-13.0$ & $21124.2$ & $6979.0$ & $186.8$ & $13119.9$ & $7037.0$ & $156.2$ & $1.12$ & $0.14$ & $1.20$ \\
   $-12.2$ & $20130.9$ & $6874.3$ & $185.1$ & $12288.4$ & $6218.9$ & $135.1$ & $1.12$ & $0.23$ & $1.37$ \\
   $-11.2$ & $19671.6$ & $6497.7$ & $148.9$ & $11886.6$ & $5407.1$ & $110.0$ & $1.12$ & $0.23$ & $1.35$ \\
   $-6.3$  & $19046.4$ & $5867.3$ & $64.3$  & $11031.6$ & $5187.0$ & $90.0$  & $1.10$ & $0.23$ & $0.71$ \\
   $-4.4$  & $18436.0$ & $6681.7$ & $58.9$  & $10732.1$ & $5103.7$ & $96.0$  & $1.05$ & $0.15$ & $0.61$ \\
   $+2.7$  & $16578.8$ & $5091.3$ & $37.6$  & $10370.4$ & $6877.8$ & $163.0$ & $1.19$ & $0.20$ & $0.23$ \\  
   \hline
 \end{tabular}
\label{caii_NIR_gauss_tab}
\end{table*}

The \ion{Ca}{ii} H\&K region is more complex due to additional
photospheric \ion{Si}{ii} at $\lambda 3856$, coupled with blanketing
from iron-peak elements that makes the definition of the
pseudo-continuum less reliable. Following \citet{maguire14} and
\citet{silverman15}, we modelled each pair of \ion{Ca}{ii} H\&K lines
using two Gaussians where the relative positions and their widths were
fixed, and the relative strength of the two Gaussians fixed to unity.
We consider the HV and photospheric \ion{Ca}{ii} components, the
\ion{Si}{ii} $\lambda3856$ line, and a weak feature at
$\sim$3600\,\AA, possibly caused by the blending of several lines of
iron-peak elements (the feature can be seen at $\simeq$27000\kms\ in
Fig.~\ref{caii_velplot_fig}).  We fixed the position of the
\ion{Si}{ii} $\lambda 3856$ line to be within 5 per cent of the
photospheric velocity of the \ion{Ca}{ii} NIR line, and the
velocity of the \ion{Ca}{ii} $H\&K$ HV and photospheric components
to be within 10 per cent of the values from the \ion{Ca}{ii} NIR lines.

The \ion{Si}{ii} feature is required in our fits from $-11$\,d.  The
weak feature at $\sim$3600\,\AA\ is blended with the \ion{Ca}{ii} H\&K
line at early stages, but around maximum light the feature becomes
more detached, becoming an independent feature on the blue side of the
H\&K profile around peak brightness. 
The extension of the \ion{Ca}{ii} material reaches a velocity of
$\sim$29000\kms\ in the very outermost layers of SN\,2015F.

\subsubsection{Velocity evolution of Ca\,II}
\label{sec:veloc-evol-ca}

\begin{figure}
\includegraphics[width=90mm]{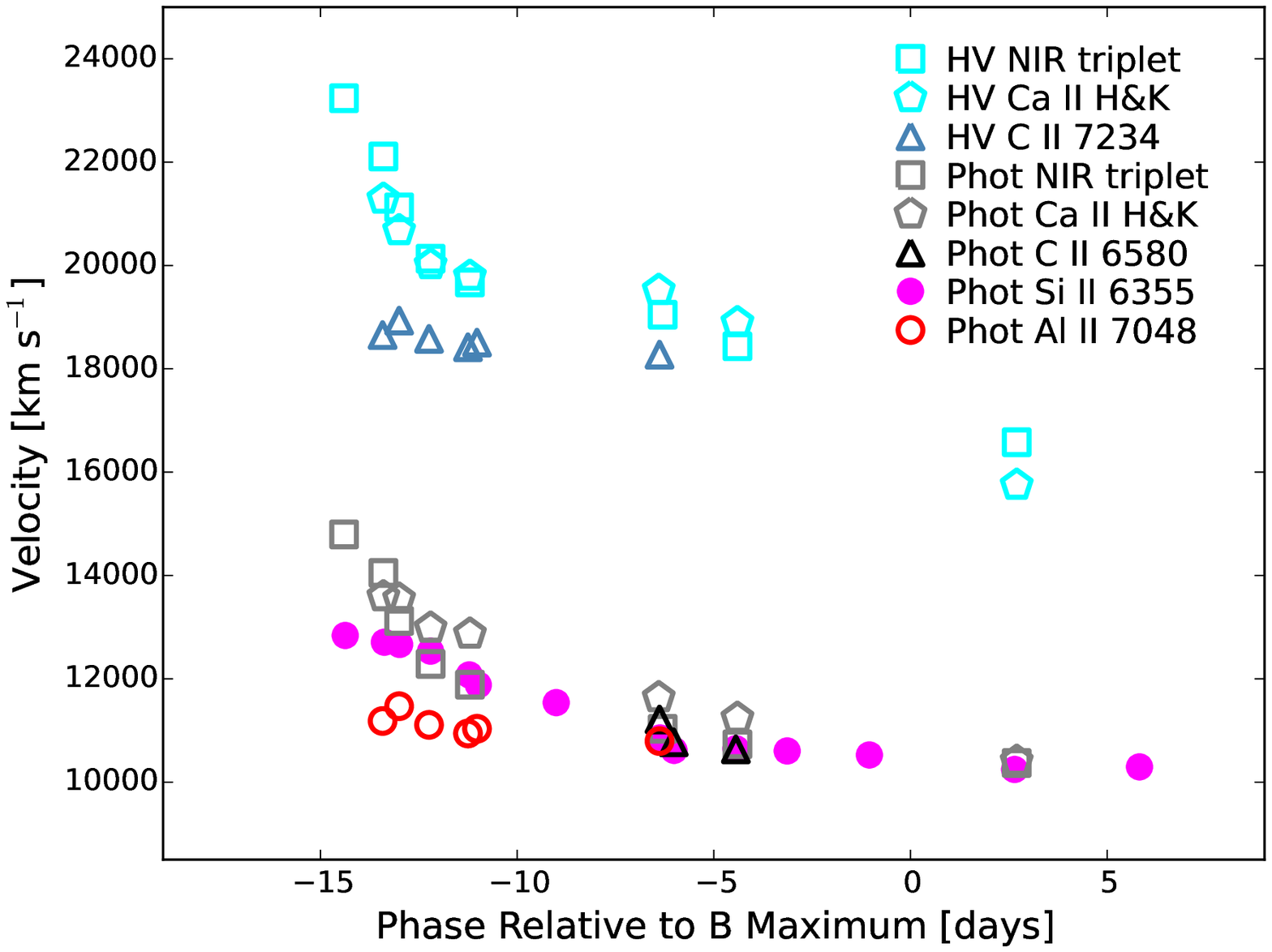}
\caption{The expansion velocities for different ions in SN\,2015F as a
  function of phase: HV \ion{Ca}{ii} NIR triplet (cyan squares), HV
  \ion{Ca}{ii} H\&K (cyan pentagons), photospheric \ion{Ca}{ii} NIR
  triplet (grey squares), photospheric \ion{Ca}{ii} H\&K (grey
  pentagons), \ion{Si}{ii} $\lambda 6355$ (magenta dots), and
  photospheric \ion{C}{ii} $\lambda6580$ (black triangles). We also
  show the 6800\,\AA\ line intepreted as either photospheric
  \ion{Al}{ii} (open circles) and HV \ion{C}{ii} (blue triangles).}

\label{ions_velplot_fig}
\end{figure}

Fig.~\ref{ions_velplot_fig} shows the expansion velocity of the
\ion{Ca}{ii} lines as function of phase. The median difference between
the HV and the photospheric components for the NIR and H\&K features
is $7850$\kms\ and $7350$\kms, respectively, consistent with previous
studies using larger samples \citep{maguire14, silverman15}. The HV
\ion{Ca}{ii} component shows a dramatic velocity evolution over
$-14$\,d to $-11$\,d, evolving from 23000\kms\ to 19500\kms\ in three
days. It then plateaus, decreasing only $\sim500$\kms\ over the next
five days.  This plateau is coincident with a transition from equally
strong HV and photospheric components, to a dominant photospheric
component (see Table~\ref{caii_NIR_gauss_tab}).  After $-6$\,d, the
velocity of the HV \ion{Ca}{ii} again declines more rapidly.

We find a very good agreement between the \ion{Si}{ii} $\lambda 6355$
velocity (measured from the minimum of the feature) and the
\ion{Ca}{ii} photospheric velocities (measured with the Gaussian
fitting), remarkable as the techniques are quite independent. Only at
epochs prior to $-11$\,d do we see a departure in the behaviour of the
\ion{Si}{ii} velocity from the \ion{Ca}{ii} photospheric velocity,
with the \ion{Si}{ii} showing a shallower evolution; this may be the
result of contamination from HV \ion{C}{ii} $\lambda 6580$ moving the
minimum of the line profile to redder wavelengths.

Finally, we see a different behaviour in the velocity evolution of the
HV \ion{Ca}{ii} and photospheric lines, and the 6800\,\AA\ absorption
feature (Fig.~\ref{ions_velplot_fig})
While the other lines show a consistent evolution, the 6800\,\AA\
feature shows a evolution similar to a plateau with a small slope.
We note that the velocity measurement at $-6$\,d is hampered by a
nearby telluric line, and at later epochs the minimum of the feature
is strongly affected by telluric absorption.

\subsubsection{Strength of the Ca\,II features}
\label{sec:strength-ca-feat}

\citet{childress14} defined $R_\mathrm{HVF}$ as the ratio between the
pEWs of the HV and photospheric \ion{Ca}{ii} NIR components.  In
Table~\ref{caii_NIR_gauss_tab} we list the evolution of
$R_\mathrm{HVF}$ as function of phase. The HV component is dominant
over the photospheric feature from $-13$ to $-11$\,d, and then
declines in strength. At phases later than $-10$ to $-7$\,d, the
`photospheric' \ion{Ca}{ii} component begins to dominate.  Close to
maximum light, SN\,2015F has $R_\mathrm{HVF}\simeq0.23$ consistent
with the results of \citet{childress14} and \citet{maguire14}.  In
particular, the $R_\mathrm{HVF}$ value for SN\,2015F, and its $\Delta
m_{15}(B)=1.35$, perfectly fits in figure 2 of \citet{childress14}.

\subsection{C\,II in SN\,2015F}
\label{carbon_sec}

\begin{figure*}
\includegraphics[width=150mm]{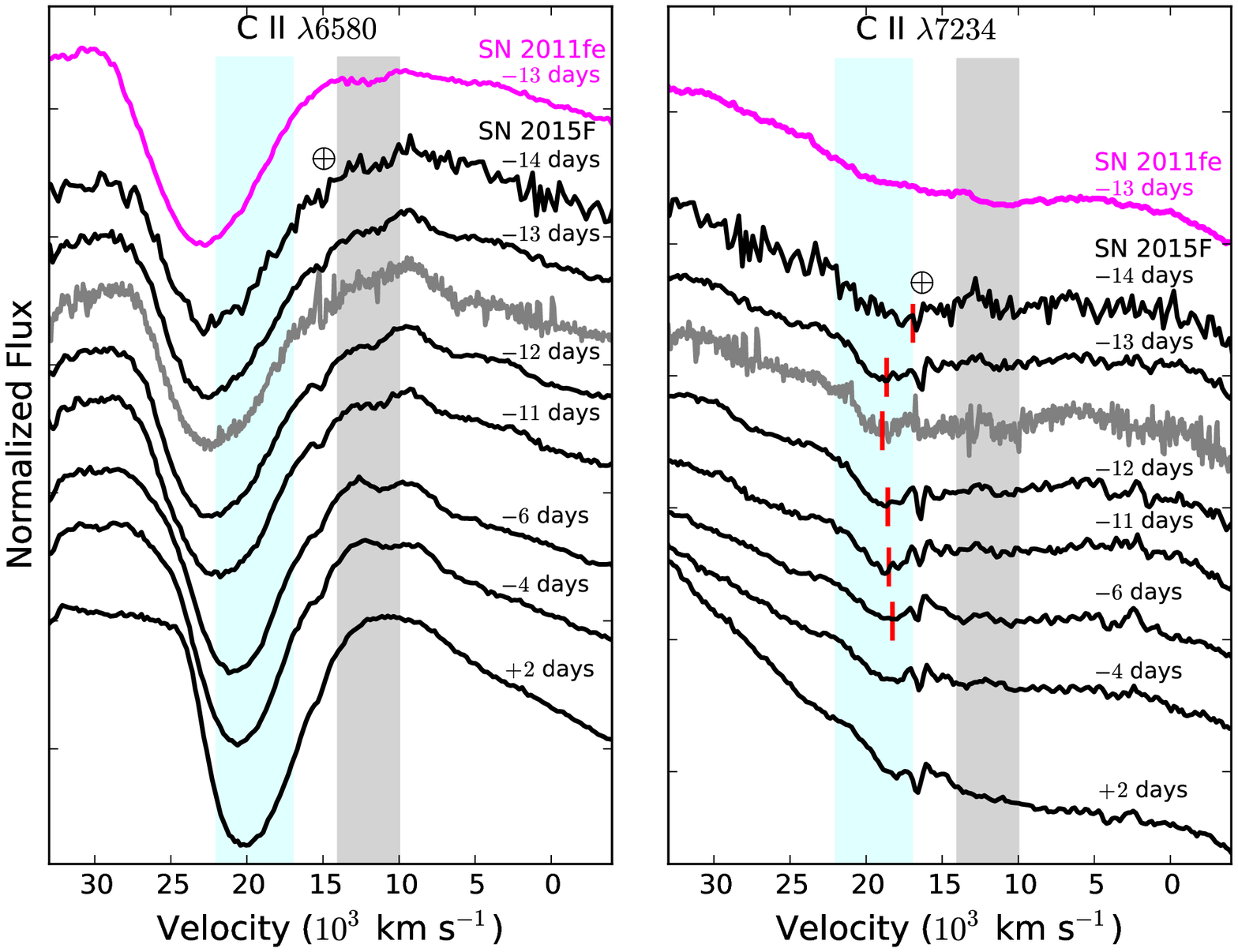}
\caption{The spectral region around \ion{C}{ii} $\lambda 6580$ (left
  panel) and $\lambda 7234$ (right panel) in velocity space. The
    spectra obtained with EFOSC/NTT are shown in black, and in grey
    the  WiFeS/ANU 2.3-m spectrum at $-13$\,d. In the right panel,
  the red vertical lines mark the velocity measured from the minimum
  of the 6800 \AA\ feature. For comparison, we show the $-13$\,d
  spectrum of SN\,2011fe (magenta).  The grey and cyan regions mark
  the position of the \ion{C}{ii} lines moving at
  $10000$--$14000$\kms\ (photospheric) and at $17000$--$22000$\kms\
  (HV).}
\label{cii_velplot_fig}
\end{figure*}

In Fig.~\ref{cii_velplot_fig}, we present the spectral sequence of
SN\,2015F around the \ion{C}{ii} $\lambda6580$ and $\lambda7234$ lines
in velocity space, compared to SN\,2011fe. The \ion{C}{ii} $\lambda
6580$ photospheric absorption is detected until $-4$\,d at velocities
similar to \ion{Si}{ii} (see Fig.~\ref{ions_velplot_fig}), although the
measurement is difficult at phases prior to $-10$\,d as the feature
does not have a well defined minimum; see Fig.~\ref{cii_velplot_fig}).
We also note that around maximum light, the \ion{Si}{ii} $\lambda
6355$ line becomes stronger while the \ion{C}{ii} lines become weaker;
thus P-Cygni emission from \ion{Si}{ii} may affect the \ion{C}{ii}
feature by moving the minimum to redder wavelengths (lower
velocities).

Fig.~\ref{cii_velplot_fig} also presents a spectral sequence showing
the evolution of the $\sim 6800$\,\AA\ absorption feature. One
interpretation for this feature is that it is HV \ion{C}{ii},
which we discuss in Section \ref{sec:carbon-material}.

\section{Spectral Modelling}
\label{spec_mod_sec}

We next investigate whether simple spectral modelling can assist with
further identification of the lines in the spectra of SN\,2015F.  We
used \textsc{syn++} \citep{thomas11a}, an updated version of
\textsc{synow} \citep{fisher97}, to model the spectra. The physical
assumptions of \textsc{syn++} match those of \textsc{synow}
\citep{2000PhDT.........6F}, so our findings are restricted to the
identification of features and not quantitative abundances.

\begin{figure*}
\includegraphics[width=190mm]{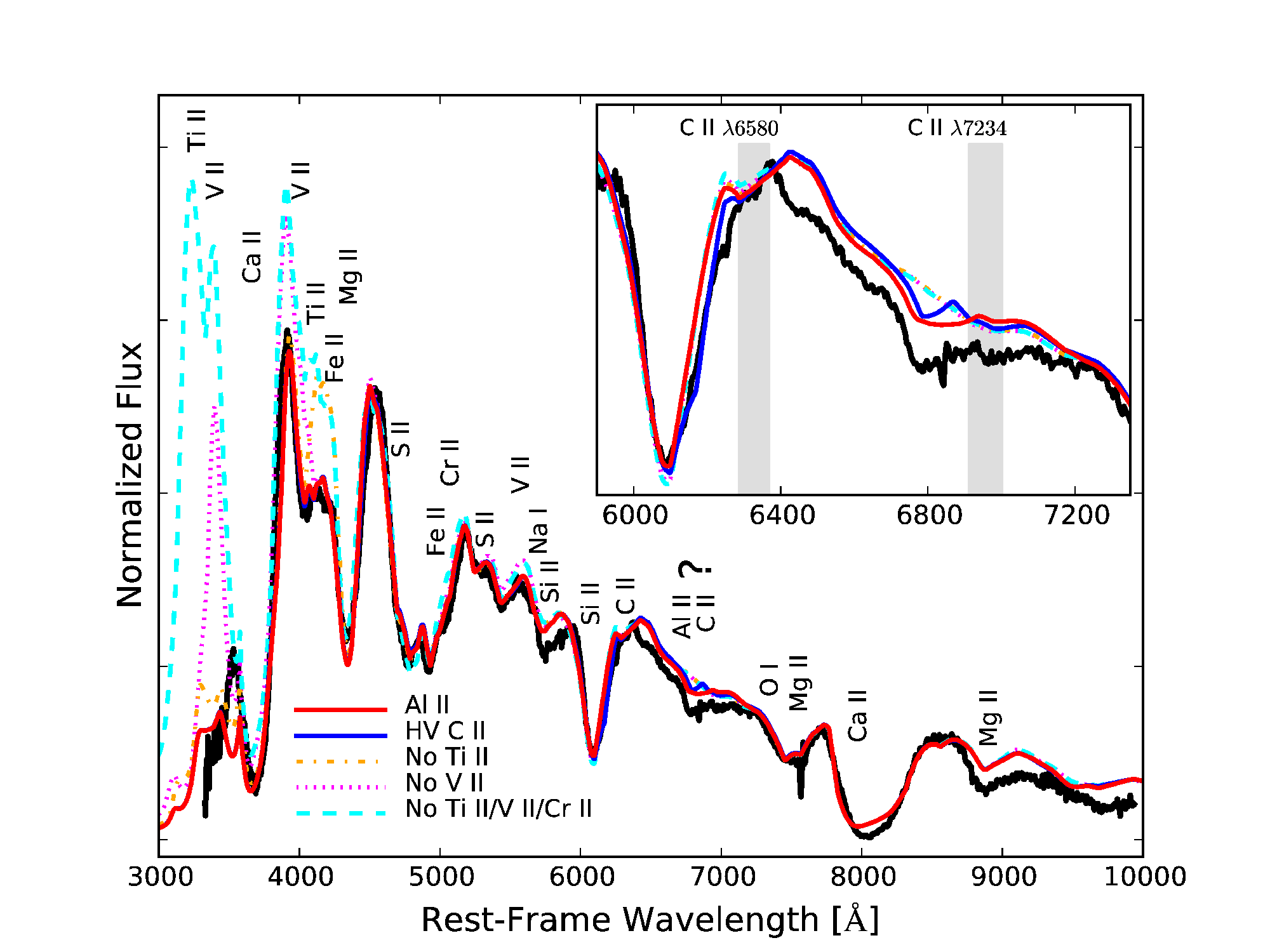}
\caption{The $-13$\,d spectrum of SN\,2015F (black line) along with
  various \textsc{syn++} models (see Section~\ref{spec_mod_sec} for
  details). The red line corresponds to a \textsc{syn++} model that
  includes photospheric \ion{Al}{ii} expanding at a velocity of
  $13000$\kms, and iron-peak elements (\ion{V}{ii}, \ion{Ti}{ii}, \ion{Cr}{ii},
  and \ion{Fe}{ii}) expanding at a velocity of $14800$\kms. The blue
  line corresponds to a similar base model, but now includes a HV \ion{C}{ii} component
  expanding at a velocity of $20000$\kms, and not \ion{Al}{ii}.
  To show the contribution of iron-peak elements, we present the same
  base model but computed excluding \ion{Ti}{ii} (orange dot-dashed line), 
  \ion{V}{ii} (magenta dotted line), and excluding \ion{Ti}{ii}, \ion{V}{ii},
  and \ion{Cr}{ii} (cyan dashed line). In these models we do not include HV
  \ion{C}{ii} or \ion{Al}{ii}. The inset shows the region around the \ion{C}{ii}
  lines and the $\sim 6800$\AA\ feature. The grey regions mark the position
  of the photospheric \ion{C}{ii} lines moving at $10000$--$14000$\kms.
  The ions responsible for prominent spectral features are indicated on the figure.
  The spectrum of SN\,2015F is corrected by Milky Way and host galaxy reddening.}
\label{model_m13_fig}
\end{figure*}

In Fig.~\ref{model_m13_fig} we show the spectrum of SN\,2015F at
$-13$\,d along with a \textsc{syn++} model that reproduces the main
spectroscopic features reasonably well.  We assumed a photosphere
expanding at a velocity of $12000$\kms\ and a black body temperature
of $9500$\,K, and allowed various individual ions to be detatched from
the photosphere; Table~\ref{syn++_params_tab} lists the velocities and
e-folding length scales of the main ions in our models at three
representative phases; -13\,d, -11\,d and -4\,d. We note that to
reproduce the -13\,d spectrum over the region from 4000 to 5000\,\AA,
we required \ion{V}{ii}, \ion{Ti}{ii}, \ion{Cr}{ii} and a significant
amount of \ion{Fe}{ii} at $14800$\kms. The lines of these iron group
elements are therefore produced at a significantly higher velocity
than the photosphere, and the e-folding length-scale of \ion{Fe}{ii}
is relatively large ($2300$\kms). This suggests that iron-group
material extends well beyond the photosphere, possibly reaching
expansion velocities of $\sim20000$\kms. In Fig.~\ref{model_m13_fig},
we also present various \textsc{syn++} models with different
combinations of the iron group lines to show the contribution of these
ions in the final model.

\begin{figure*}
\includegraphics[width=190mm]{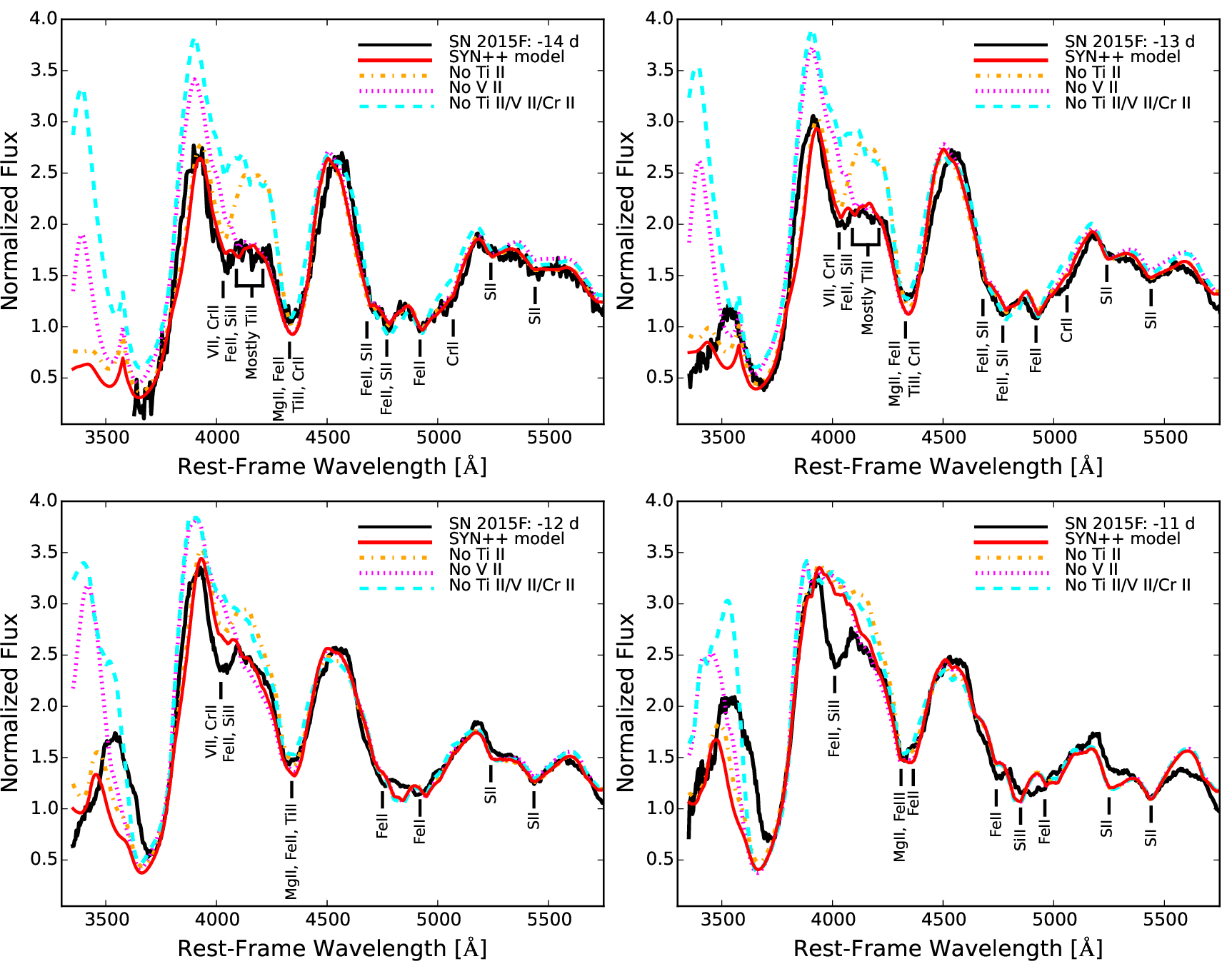}
\caption{Early time spectra of SN\,2015F (black line) along with
  various \textsc{syn++} models (see Section~\ref{spec_mod_sec} for
  details). The red line corresponds to our best \textsc{syn++} model
  that includes iron-peak elements (\ion{V}{ii}, \ion{Ti}{ii}, \ion{Cr}{ii}
  and \ion{Fe}{ii}). To show the contribution of iron-peak elements,
  we present the same base model but computed excluding \ion{Ti}{ii}
  (orange dot-dashed line), \ion{V}{ii} (magenta dotted line), and \ion{Ti}{ii},
  \ion{V}{ii}, and \ion{Cr}{ii} (cyan dashed line). In these models we do
  not include HV \ion{C}{ii} or \ion{Al}{ii}. The ions responsible for
  prominent spectral features are indicated on the figure.}
\label{model_iron_peak_fig}
\end{figure*}

In Fig.~\ref{model_iron_peak_fig} we overplot our \textsc{syn++}
models on top of the first four spectra of SN\,2015F, and focus on the
blue part of the spectra, which are dominated by lines of iron-peak
elements. At -14\,d and -13\,d, the features are very broad, and the
spectra exhibit strong absorption features from 3900--4500\,\AA\
mainly produced by iron-peak elements.  At -12\,d the strength of the
\ion{V}{ii}, \ion{Ti}{ii} and \ion{Cr}{ii} lines begin to decrease.
At -11\,d the line profiles become narrower, and the features produced
by \ion{V}{ii}, \ion{Ti}{ii} and \ion{Cr}{ii} are no longer clear.
Lines of \ion{Fe}{iii}, and possibly \ion{Si}{iii}, begin to appear.

The strong absorption at $\sim$6800\,\AA\ can be reproduced by a HV
\ion{C}{ii} component at 20000\kms, or by photospheric \ion{Al}{ii} at
a velocity of 13000\kms. For the HV \ion{C}{ii} component, we adjusted
$T_{\mathrm{ext}}=14000$\,K to make the HV \ion{C}{ii} $\lambda 7234$
line stronger than the HV \ion{C}{ii} $\lambda 6580$ line, thus
mimicking as closely as possible the profile in the SN\,2015F
spectrum. However, as for the $\mathcal{R}(\ion{Si}{ii})$ ratio,
non-LTE effects may play a role \citep[see][]{nugent95, hachinger08}.
These are not captured by the LTE assumption of \textsc{syn++}, and
therefore the $T_{\mathrm{ext}}$ used for the HV \ion{C}{ii} component
might not be a reliable estimation of the true value. Even after this
adjustment in $T_{\mathrm{ext}}$, the HV \ion{C}{ii} only partially
reproduces the strong absorption feature seen at $\sim$6800\,\AA. A
stronger HV component of \ion{C}{ii} $\lambda 7234$ and \ion{C}{ii}
$\lambda4267$ may yield a better model of the $\sim$6800\,\AA\ and the
$\sim$4030\,\AA\ absorption features, respectively.

A combination of a HV \ion{C}{ii} component and of \ion{Al}{ii} line
is also a possibility to explain the $\sim 6800$~\AA\ feature, but in terms
of \textsc{syn++} modelling difficult to disentangle since it yields
similar results to the models displayed in Fig.~\ref{model_m13_fig}.

\begin{table*}
 \caption{The \textsc{syn++} parameters for the spectral fits at $-13$\,d, $-11$\,d, and $-4$\,d.}
 \label{symbols}
 \begin{tabular}{@{}lccccccccc}
  \hline
 Ion  & \multicolumn{3}{c}{$-13$\,d model}                &  \multicolumn{3}{c}{$-11$\,d model}               & \multicolumn{3}{c}{$-4$\,d model} \\
      & $v_{\mathrm{min}}$ & $T_{\mathrm{ext}}$ & $v_{e}$ & $v_{\mathrm{min}}$ & $T_{\mathrm{ext}}$ & $v_{e}$ & $v_{\mathrm{min}}$ & $T_{\mathrm{ext}}$ & $v_{e}$ \\
      & [$10^{3}$\kms] & [$10^{3}$\,K] & [$10^{3}$\kms]   & [$10^{3}$\kms] & [$10^{3}$\,K] & [$10^{3}$\kms]   & [$10^{3}$\kms] & [$10^{3}$\,K] & [$10^{3}$\kms]   \\
\hline
Photosphere & $12.0$ & $9.5$ & ...     & $11.5$ & $10.0$ & ...     & $10.0$ & $11.5$ & ...   \\
C~II        & $14.0$ & $9.5$ & $1.30$  & $13.5$ & $10.0$ & $1.30$  & $11.5$ & $11.5$ & $1.0$ \\
O~I         & $13.0$ & $9.5$ & $2.50$  & $12.5$ & $10.0$ & $1.50$  & $12.5$ & $11.5$ & $1.0$ \\
Na~I        & $12.0$ & $9.5$ & $1.00$  & $11.5$ & $10.0$ & $1.00$  & $10.0$ & $11.5$ & $1.0$ \\
Mg~II       & $13.0$ & $9.5$ & $1.50$  & $12.5$ & $10.0$ & $1.50$  & $12.5$ & $11.5$ & $1.0$ \\
Si~II       & $13.6$ & $9.5$ & $1.90$  & $12.9$ & $10.0$ & $1.80$  & $11.5$ & $11.5$ & $1.7$ \\
Si~III      & ...    & ...   & ...     & ...    & ...    & ...     & $10.0$ & $11.5$ & $1.0$ \\
S~II        & $12.1$ & $9.5$ & $1.35$  & $11.7$ & $10.0$ & $1.30$  & $11.0$ & $11.5$ & $1.0$ \\ 
Ca~II       & $13.0$ & $9.0$ & $4.00$  & $13.3$ & $10.0$ & $2.00$  & $11.0$ & $11.5$ & $1.5$ \\
HV Ca~II    & ...    & ...   & ...     & $20.0$ & $10.0$ & $3.00$  & $18.5$ & $11.5$ & $1.5$ \\
Ti~II       & $14.8$ & $9.5$ & $1.80$  & $13.5$ & $10.0$ & $1.50$  & ...    & ...    & ...   \\
V~II        & $14.8$ & $9.5$ & $2.00$  & $14.8$ & $10.0$ & $2.00$  & ...    & ...    & ...   \\
Cr~II       & $14.8$ & $9.5$ & $2.00$  & ...    & ...    & ...     & ...    & ...    & ...   \\
Fe~II       & $14.8$ & $9.5$ & $2.30$  & $12.5$ & $10.0$ & $1.50$  & $11.5$ & $11.5$ & $2.0$ \\
Fe~III      & ...    & ...   & ...     & $12.0$ & $10.0$ & $1.50$  & $10.0$ & $11.5$ & $1.0$ \\
\hline                                                                                                              
Al~II       & $13.0$ & $9.5$  & $1.50$ & $12.3$ & $10.0$ & $1.50$  & $12.0$ & $11.5$ & $1.0$ \\
HV C~II     & $20.0$ & $14.0$ & $2.10$ & $20.0$ & $14.0$ & $2.10$  & $19.6$ & $14.0$ & $2.0$ \\
  \hline
 \end{tabular}
\label{syn++_params_tab}
\end{table*}

\section{Discussion}
\label{discusion_sec}

The previous sections have presented a high-quality time-series of
spectra and photometry of the nearby type Ia SN\,2015F. Our data make
it one of the best observed SNe Ia at early times, and the early
spectroscopic coverage have allowed us to study the outer layers of
the SN ejecta in detail. In particular, these data provide evidence
for either photospheric \ion{Al}{ii} or high-velocity \ion{C}{ii}, as
well as iron-peak elements in the outer layers. We discuss these in
turn.

\subsection{Photospheric Aluminium}
\label{sec:aluminium}

Our favoured explanation of the $\sim6800$\,\AA\ spectral feature in
SN\,2015F is photospheric \ion{Al}{ii} (see
Section~\ref{spec_mod_sec}), expanding at a velocity of $\sim
13000$\kms\ (Fig.~\ref{ions_velplot_fig}).  The \ion{Al}{ii} material
has to be confined in a relatively narrow range of velocity, as the
$\sim 6800$\,\AA\ feature does not appear to evolve in velocity over
16 days (Fig.~\ref{cii_velplot_fig}).  However, we caution that the
feature is quite weak and is affected by telluric absorption; a
definitive statement about the velocity evolution is difficult to
make.

Aluminium in SNe Ia has not been commonly reported in the literature.
To our knowledge, the only previous claim was in the peculiar `.Ia'
\citep{2007ApJ...662L..95B} candiate SN\,2010X \citep{kasliwal10}.
$^{27}$\ion{Al}{} is the only stable aluminium isotope, which
according to nucleosynthesis calculations is $\sim 10^{3}$ more
abundant than the radioactive $^{26}$\ion{Al}{} isotope
\citep{iwamoto99,seitenzahl13}. However, the expected mass fraction of
$^{27}$\ion{Al}{} in SNe Ia is relatively low, only $10^{-3}$ to
$10^{-2}$ times the total mass of $^{28}$\ion{Si}{}
\citep{iwamoto99,seitenzahl13}, the latter being the most abundant
silicon isotope in SNe Ia.

Given this low predicted abundance of Al, strong \ion{Al}{ii} features
seem unexpected. The yield of $^{27}$Al obtained from the W7
\citep{nomoto84} nucleosyntheis models of \citet{iwamoto99}, and the
three-dimensional N100 delayed-detonation models of
\citet{seitenzahl13}, predict a strong dependence of the abundance of
$^{27}$\ion{Al}{} on metallicity. A change from zero to solar
metallicity in the progenitor white dwarf produces an increase of an
order of magnitude in the yield of $^{27}$Al by mass. As a comparison,
the abundances of $^{12}$C and $^{28}$Si remain essentially flat as
function of progenitor metallicity.  Thus a relatively metal-rich
progenitor may help to explain the presence of Al in SN\,2015F.

  In Section \ref{sec:comp-other-supern} we noted the common
  presence of the $\sim6800$\,\AA\ feature in SN\,1991bg-like SNe.
  This class appears to explode in more massive, higher metallicity
  galaxies; we also note the non-detection of this feature in SN\,2011fe
  which seems to be the result of a sub-solar metallicity progenitor
  \citep[see][]{mazzali14}.  As SN\,1991bg-like SNe exhibit lower
  photospheric temperatures than normal SNe Ia, in principle, the
  presence of the \ion{Al}{ii} lines could be explained by a
  temperature effect and not as a metallicity effect. However, in the case of
  SN\,2015F a temperature effect can be ruled out by the simultaneous
  detection of the $\sim6800$\,\AA\ feature with \ion{Si}{iii} lines,
  which are a signature of a hot SN ejecta, and are strong at -6\,d and
  -4\,d (see Section \ref{sec:comparison-with-sn}).

\subsection{Carbon material}
\label{sec:carbon-material}

A second explanation for the $\sim6800$\,\AA\ feature is high-velocity
(HV) \ion{C}{II} $\lambda 7234$; photospheric \ion{C}{II} is clearly
detected.  This suggests that the outermost layers
($\gtrapprox$18000\kms) of SN\,2015F are mostly unburned, consistent
with the \citet{mazzali14} model for SN\,2011fe, in which the
outermost layers of the SN ejecta (\textgreater$19400$\kms) are
unburned, and are composed mainly of carbon. The fact that SN\,2015F
has a faster decline rate than SN\,2011fe, and is thus a dimmer/cooler
event, suggests a less efficient burning, and perhaps an even larger
amount of unburned material in the outer layers than in SN\,2011fe.
 

In recent delayed-detonation models \citep{seitenzahl13},
the outermost layers of the ejecta ($v_{exp}$ \textgreater $20000$ \kms)
are mostly composed of carbon and oxygen, and this may
explain any HV \ion{C}{ii}. Nevertheless, we do not see a correspondingly
strong HV \ion{O}{i} line in SN\,2015F. In \citet{seitenzahl13},
carbon could also be present down to about $10000$\kms, which may explain
photospheric \ion{C}{ii} in SN\,2015F \citep[but see also][]{mazzali14}.

Under the assumption that the 6800\,\AA\ absorption feature
corresponds to HV \ion{C}{ii} $\lambda 7234$, we show its velocity
evolution in Figs.~\ref{ions_velplot_fig} and \ref{cii_velplot_fig}.
The first spectrum of SN\,2015F has lower signal-to-noise ratio
implying a larger uncertainty in the minimum of the feature, located
at $\simeq$16900\kms\ ($6828$\,\AA).  The feature evolves getting
weaker and moving to redder wavelengths with time. We measured the
minimum of this absorption feature at phases $<-6$\,d (note the
measurement is affected by telluric on the red side; see
Fig.~\ref{cii_velplot_fig}), and we show its expansion velocity in
Fig.~\ref{ions_velplot_fig}.  The feature appears confined to a narrow
range in velocity space from $\simeq$18700\kms\ to $\simeq$17000\kms,
but is persistent, and is still present in the spectrum at $+2$\,d.
The feature is also observed in SN\,2007af at slightly higher
velocities until $-4$\,d, and then disappears.

The possibility that the HV features of \ion{C}{II} and \ion{Ca}{II}
are produced close in velocity space may suggest a common origin for
the HV material (see Fig.~\ref{ions_velplot_fig}). HV \ion{Ca}{ii}
features exhibit a plateau in their velocity evolution between $-10$
and $-4$\,d (Fig.~\ref{ions_velplot_fig}). At the same phase, the
velocity measured for the possible HV \ion{C}{II} $\lambda 7234$
feature is similar, but slightly lower.

\subsection{Iron-group elements in SN\,2015F}
\label{sec:iron-group-elements}

Using \textsc{syn++} to model the spectra of SN\,2015F (see
Section~\ref{spec_mod_sec}), we have identified lines of \ion{Ti}{ii},
\ion{V}{ii}, \ion{Cr}{ii}, and \ion{Fe}{II} expanding at
$\simeq$14800\kms. This implies a non-negligible amount of
iron-group elements in the region between $15000$ to $20000$\kms
of the SN ejecta. \citet{hatano99} reported strong \ion{Fe}{ii} absorptions
in the $-12$\,d spectrum of SN\,1994D at $\sim$4300 and $\sim$4700\,\AA,
and included in their \textsc{synow} model a HV \ion{Fe}{II} component extending from
$22000$\kms\ to $29000$\kms, to reproduce these features.

Strong and broad lines of iron-peak elements at such high-expansion
velocities are generally unexpected in SNe Ia, since the pre-expansion
suffered by the layers at higher velocities than $10000$ to $13000$ \kms\
will decrease the density too much to burn the material to iron-peak
elements. Only in the case of a rapid transition in the burning speed front,
from a sub-sonic deflagration to a super-sonic detonation, would the flame burn
the outermost layers ($v_{\mathrm{exp}}$ \textgreater $13000$ \kms) to iron-peak
elements, yielding mainly radioactive $^{56}$Ni and not enough IMEs to reproduce
the characteristic spectral features of normal SNe Ia. The decay of $^{56}$Ni
mixed in the outermost layers can then heat the ejecta, producing strong lines
of doubly-ionized species such as \ion{Fe}{iii} and \ion{Si}{iii} as in the
brightest SNe. In the earliest spectra of SN\,2015F ($<-12$\,d),
this is not observed; by contrast, SN\,2015F exhibits a spectrum dominated
mainly by singly-ionized species (\ion{Fe}{ii}, \ion{Si}{ii}, \ion{S}{ii},
\ion{Ca}{ii}), consistent with a normal or relatively low ejecta temperature.
At later phases, from about $-11$\,d, SN\,2015F begins to exhibit
\ion{Si}{iii} $\lambda 4560$ and \ion{Si}{iii} $\lambda 5540$ lines,
now suggesting heating from the decay of radioactive material mixed in
the outer layers of the SN ejecta.

The absorption features produced by iron-group elements in the very
early spectra of SNe Ia could be explained by iron-peak elements
synthesized during the SN explosion and mixed to the outermost layers
of the SN ejecta, or as absorptions of iron-peak elements present in
the white dwarf surface at the moment of the explosion \citep[see][]{lentz00}
-- or a combination of both. Recent three-dimensional delayed detonation models
predict that freshly synthesized iron-peak elements are located mainly at intermediate
velocities \citep[$\sim$3000 to 10000\kms;][]{seitenzahl13}, the latter
not corresponding with the observations of SN\,2015F. However,
\citet{seitenzahl13} remark that models with a strong (turbulent)
deflagration phase, which are rather symmetric under rotation on
large scales, exhibit strong inhomogeneities in the burning products
on small scales. This may explain pockets of
iron-peak material, observed at high-velocity, mixed in the outermost layers
as in the case of SN\,2015F. 



To disentangle the metallicity of the progenitor from the fraction of
freshly synthesized iron-group elements mixed to the outermost layers
would require a detailed modelling using the abundance tomography
technique \citep{stehle05, mazzali05, tanaka11, hachinger13,
  mazzali14}. This is beyond the scope of this paper, and will be the
subject of a future article. 

\section{Summary}
\label{conclusions_sec}

We have presented spectroscopic and photometric data for the nearby
type Ia supernova SN\,2015F, obtained as part of PESSTO. In particular,

\begin{enumerate}

\item We show that SN\,2015F is a normal, low-velocity gradient (LVG)
  SN Ia. The values of the parameters $\mathcal{R}(\ion{Si}{ii})$
  (Section~\ref{sec:expans-veloc}) and $R_{\mathrm{HVF}}$
  (Section~\ref{sec:strength-ca-feat}) are consistent with its decline
  rate ($\dmB=1.35 \pm 0.03$).

\item We find moderate host galaxy reddening of
  $E(B-V)_{\mathrm{host}} = 0.085 \pm 0.019$\,mag. Assuming $H_{0} =
  70 \pm 3.0$\,\kmsmpc, and the decline rate/peak luminosity
  calibrations of \citet{phillips99} and \citet{kattner12}, we
  estimate $\mu_{\mathrm{optical}} = 31.64\pm0.14$, and
  $\mu_{\mathrm{NIR}}=31.68\pm0.11$.

\item We model the Ca\,\textsc{ii} H\&K and NIR triplet profiles to
  estimate the expansion velocity and pseudo equivalent-widths of the
  photospheric and high-velocity componets. We find that the
  high-velocity Ca\,\textsc{ii} reached an expansion velocity of
  $\simeq$23000\kms, decreasing to $\simeq$16500\kms\ just after
  maximum brightness.  The expansion velocity of the photospheric
  Ca\,\textsc{ii} component ranges from $\sim$14500\kms\ at $-14$\,d
  to $\sim$10300\kms\ at maximum light.

\item We identify photospheric \ion{C}{ii} material moving at
  $\simeq$14000\kms\ at the earliest epochs, which remains detectable
  until $-4$\,d at an expansion velocity of $\simeq$11000\kms.

\item We identify a broad absorption feature at $\sim$6800\,\AA,
  previously unremarked upon in normal SNe Ia. We offer two possible
  explanations for this feature (Section~\ref{spec_mod_sec}): our
  favoured scenario is that it is produced by photopsheric
  \ion{Al}{ii} $\lambda 7054$ expanding at $13000$-$11000$\kms. An
  overabundance of \ion{Al}{ii} relative to other SNe could be the
  result of a relatively high metallicity of the progenitor (Section
  \ref{sec:aluminium}). An alternative scenario is that it is produced
  by high-velocity (HV) \ion{C}{ii} $\lambda 7234$ expanding at
  $20000$-$18000$\kms\ (see Section \ref{sec:carbon-material}). The
  $\sim$6800\,\AA\ feature is also common in SN\,1991bg-like SNe.

\item We use \textsc{syn++}, to model the spectra of SN\,2015F.
  We find lines of Fe-peak elements such as \ion{Ti}{ii},
  \ion{V}{ii}, \ion{Cr}{ii}, and \ion{Fe}{ii} expanding at
  high velocity ($>14800$\kms) in the outermost layers of the SN.
  The inclusion of \ion{V}{ii} improves significantly our
  \textsc{syn++} models at early times (-14\,d and -13\,d).

\end{enumerate}

\section*{Acknowledgments}
We thank to the anonymous referee for his careful review that helped
to improve this manuscript.
We acknowledge support from STFC grant ST/L000679/1 and EU/FP7-ERC
grant No. [615929]. Support for G.P. is provided by the Ministry
of Economy, Development, and Tourism's Millennium Science Initiative
through grant IC120009, awarded to The Millennium Institute of
Astrophysics, MAS. K.M. acknowledges support from the STFC through
an Ernest Rutherford Fellowship. A.G.-Y. is supported by the EU/FP7
via ERC grant No. [307260], the Quantum Universe I-Core program by
the Israeli Committee for Planning and Budgeting and the ISF; by Minerva
and ISF grants; by the Weizmann-UK ``making connections'' program;
and by Kimmel and YeS awards. S.J.S. acknowledges funding from ERC Grant
agreement No. [291222] and STFC grants ST/I001123/1 and ST/L000709/1.
This material is based upon work supported by the National Science
Foundation under Grant No. 1313484.
We are grateful to Bruno Leibundgut and Masayuki
Yamanaka for providing the spectra of SN~1990N and SN~2012ht,
respectively.  This work is based on observations collected at the
European Organisation for Astronomical Research in the Southern
Hemisphere, Chile as part of PESSTO, (the Public ESO Spectroscopic
Survey for Transient Objects Survey) ESO programme 191.D-0935, and 
on observations obtained with the Southern African Large Telescope
(SALT) under program 2014-2-SCI-070.

\bibliographystyle{mnras}
\bibliography{references} 

\appendix

\section{Photometric sequence and photometry of SN\,2015F}

\begin{table*}
\caption{$UBVgri$ photometric sequence around SN\,2015F.}
\label{op_local_seq_tab}
\begin{tabular}{@{}cccccccc}
\hline
R.A. & Decl. & $U$ & $B$ & $V$ & $g$ & $r$ & $i$ \\
\hline
$07^{\mathrm{h}}35^{\mathrm{m}}04\fs6$ & $-69\degr28\arcmin07\farcs9$ & 17.348(0.036) & 17.086(0.021) & 16.245(0.015) & 16.624(0.015) & 15.974(0.020) & 15.648(0.016) \\
$07^{\mathrm{h}}35^{\mathrm{m}}05\fs6$ & $-69\degr32\arcmin42\farcs3$ & --            & 17.431(0.019) & 16.119(0.015) & 16.753(0.015) & 15.627(0.034) & 15.114(0.016) \\
$07^{\mathrm{h}}35^{\mathrm{m}}13\fs0$ & $-69\degr29\arcmin57\farcs6$ & 14.986(0.015) & 14.724(0.015) & 13.894(0.015) & 14.267(0.015) & 13.648(0.022) & 13.374(0.015) \\
$07^{\mathrm{h}}35^{\mathrm{m}}13\fs7$ & $-69\degr27\arcmin05\farcs3$ & 16.902(0.026) & 16.672(0.019) & 15.851(0.015) & 16.215(0.015) & 15.615(0.028) & 15.303(0.020) \\
$07^{\mathrm{h}}35^{\mathrm{m}}15\fs3$ & $-69\degr27\arcmin11\farcs5$ & 17.793(0.051) & 17.455(0.015) & 16.631(0.015) & 17.006(0.015) & 16.366(0.027) & 16.078(0.024) \\
$07^{\mathrm{h}}35^{\mathrm{m}}23\fs1$ & $-69\degr35\arcmin15\farcs9$ & --            & 17.067(0.015) & 15.755(0.015) & 16.393(0.015) & 15.242(0.024) & 14.747(0.015) \\
$07^{\mathrm{h}}35^{\mathrm{m}}27\fs9$ & $-69\degr31\arcmin08\farcs2$ & 16.829(0.021) & 16.515(0.015) & 15.674(0.015) & 16.053(0.015) & 15.454(0.031) & 15.175(0.015) \\ 
$07^{\mathrm{h}}35^{\mathrm{m}}32\fs7$ & $-69\degr34\arcmin20\farcs4$ & 15.082(0.015) & 14.980(0.015) & 14.279(0.015) & 14.597(0.015) & 14.102(0.020) & 13.857(0.015) \\
$07^{\mathrm{h}}35^{\mathrm{m}}35\fs9$ & $-69\degr29\arcmin47\farcs4$ & 17.640(0.029) & 16.823(0.015) & 15.639(0.015) & 16.174(0.015) & 15.270(0.023) & 14.850(0.017) \\
$07^{\mathrm{h}}35^{\mathrm{m}}37\fs7$ & $-69\degr27\arcmin52\farcs7$ & 16.210(0.015) & 15.976(0.015) & 15.180(0.015) & 15.538(0.015) & 14.970(0.021) & 14.701(0.015) \\ 
$07^{\mathrm{h}}35^{\mathrm{m}}38\fs0$ & $-69\degr33\arcmin47\farcs9$ & 16.575(0.015) & 16.471(0.015) & 15.747(0.015) & 16.077(0.015) & 15.541(0.019) & 15.266(0.015) \\
$07^{\mathrm{h}}35^{\mathrm{m}}43\fs7$ & $-69\degr27\arcmin02\farcs1$ & 18.004(0.015) & 17.852(0.029) & 16.881(0.015) & 17.297(0.015) & 16.588(0.034) & 16.255(0.015) \\
$07^{\mathrm{h}}35^{\mathrm{m}}44\fs6$ & $-69\degr36\arcmin21\farcs3$ & 16.595(0.015) & 16.279(0.015) & 15.424(0.015) & 15.821(0.015) & 15.172(0.027) & 14.851(0.015) \\
$07^{\mathrm{h}}35^{\mathrm{m}}45\fs0$ & $-69\degr34\arcmin39\farcs3$ & 17.238(0.015) & 17.140(0.015) & 16.413(0.015) & 16.745(0.015) & 16.217(0.028) & 15.951(0.025) \\ 
$07^{\mathrm{h}}35^{\mathrm{m}}46\fs2$ & $-69\degr25\arcmin14\farcs4$ & 17.877(0.077) & 17.590(0.035) & 16.762(0.020) & 17.136(0.015) & 16.523(0.015) & 16.239(0.019) \\ 
$07^{\mathrm{h}}35^{\mathrm{m}}47\fs2$ & $-69\degr30\arcmin55\farcs2$ & 17.448(0.052) & 17.146(0.015) & 16.267(0.015) & 16.645(0.015) & 16.016(0.018) & 15.726(0.019) \\
$07^{\mathrm{h}}35^{\mathrm{m}}54\fs2$ & $-69\degr24\arcmin18\farcs9$ & 16.011(0.015) & 15.529(0.021) & 14.591(0.015) & 15.008(0.017) & 14.277(0.015) & 13.949(0.015) \\
$07^{\mathrm{h}}35^{\mathrm{m}}57\fs1$ & $-69\degr27\arcmin11\farcs7$ & 14.316(0.015) & 14.239(0.028) & 13.585(0.015) & 13.879(0.015) & 13.425(0.016) & 13.206(0.019) \\
$07^{\mathrm{h}}35^{\mathrm{m}}59\fs4$ & $-69\degr25\arcmin06\farcs6$ & 17.916(0.015) & 17.695(0.040) & 16.835(0.015) & 17.203(0.015) & 16.628(0.020) & 16.325(0.018) \\
$07^{\mathrm{h}}36^{\mathrm{m}}02\fs5$ & $-69\degr32\arcmin52\farcs7$ & 17.660(0.049) & 17.506(0.024) & 16.787(0.015) & 17.099(0.016) & 16.624(0.039) & 16.378(0.033) \\
$07^{\mathrm{h}}36^{\mathrm{m}}06\fs9$ & $-69\degr28\arcmin51\farcs4$ & 17.593(0.015) & 17.080(0.020) & 16.144(0.015) & 16.571(0.015) & 15.883(0.017) & 15.584(0.018) \\
$07^{\mathrm{h}}36^{\mathrm{m}}13\fs8$ & $-69\degr24\arcmin51\farcs1$ & 16.904(0.019) & 16.735(0.015) & 15.955(0.015) & 16.296(0.015) & 15.743(0.015) & 15.471(0.021) \\
$07^{\mathrm{h}}36^{\mathrm{m}}14\fs1$ & $-69\degr26\arcmin42\farcs1$ & 18.312(0.015) & 17.814(0.028) & 16.887(0.015) & 17.300(0.017) & 16.648(0.029) & 16.342(0.022) \\
$07^{\mathrm{h}}36^{\mathrm{m}}19\fs2$ & $-69\degr35\arcmin28\farcs7$ & --            & 18.400(0.037) & 16.915(0.015) & 17.666(0.015) & 16.306(0.023) & 15.686(0.016) \\
$07^{\mathrm{h}}36^{\mathrm{m}}23\fs3$ & $-69\degr23\arcmin58\farcs4$ & 16.278(0.032) & 15.856(0.017) & 14.960(0.015) & 15.363(0.015) & 14.690(0.017) & 14.375(0.015) \\
$07^{\mathrm{h}}36^{\mathrm{m}}24\fs7$ & $-69\degr24\arcmin57\farcs5$ & 17.479(0.024) & 16.983(0.021) & 15.925(0.015) & 16.403(0.015) & 15.604(0.019) & 15.193(0.023) \\
$07^{\mathrm{h}}36^{\mathrm{m}}26\fs3$ & $-69\degr26\arcmin17\farcs4$ & 15.398(0.015) & 15.385(0.015) & 14.775(0.015) & 15.048(0.015) & 14.633(0.015) & 14.419(0.018) \\
$07^{\mathrm{h}}36^{\mathrm{m}}28\fs8$ & $-69\degr24\arcmin58\farcs2$ & 17.419(0.023) & 17.241(0.036) & 16.424(0.015) & 16.785(0.015) & 16.188(0.015) & 15.913(0.017) \\ 
$07^{\mathrm{h}}36^{\mathrm{m}}31\fs2$ & $-69\degr29\arcmin59\farcs1$ & 17.047(0.025) & 16.957(0.036) & 16.264(0.015) & 16.575(0.015) & 16.092(0.019) & 15.852(0.042) \\
$07^{\mathrm{h}}36^{\mathrm{m}}36\fs5$ & $-69\degr36\arcmin14\farcs8$ & --            & 17.323(0.015) & 16.201(0.016) & 16.724(0.015) & 15.811(0.023) & 15.419(0.018) \\
$07^{\mathrm{h}}36^{\mathrm{m}}39\fs1$ & $-69\degr30\arcmin44\farcs3$ & 18.161(0.015) & 16.895(0.033) & 15.618(0.015) & 16.222(0.015) & 15.132(0.015) & 14.713(0.019) \\
$07^{\mathrm{h}}36^{\mathrm{m}}48\fs8$ & $-69\degr26\arcmin04\farcs2$ & 15.990(0.015) & 15.914(0.016) & 15.240(0.015) & 15.545(0.015) & 15.071(0.015) & 14.840(0.021) \\
$07^{\mathrm{h}}36^{\mathrm{m}}54\fs2$ & $-69\degr30\arcmin21\farcs3$ & 17.652(0.086) & 16.943(0.043) & 15.925(0.026) & 16.445(0.093) & 15.560(0.015) & 15.223(0.049) \\
$07^{\mathrm{h}}36^{\mathrm{m}}59\fs4$ & $-69\degr36\arcmin03\farcs3$ & 17.206(0.015) & 16.744(0.015) & 15.845(0.015) & 16.252(0.015) & 15.604(0.019) & 15.338(0.015) \\
$07^{\mathrm{h}}36^{\mathrm{m}}59\fs9$ & $-69\degr30\arcmin31\farcs0$ & 18.021(0.015) & 17.383(0.015) & 16.437(0.015) & 16.889(0.050) & 16.198(0.084) & 15.859(0.071) \\
$07^{\mathrm{h}}37^{\mathrm{m}}02\fs8$ & $-69\degr32\arcmin36\farcs0$ & 16.793(0.030) & 16.079(0.015) & 15.079(0.015) & 15.535(0.015) & 14.762(0.019) & 14.447(0.015) \\
$07^{\mathrm{h}}37^{\mathrm{m}}07\fs5$ & $-69\degr33\arcmin46\farcs2$ & 16.625(0.118) & 16.431(0.015) & 15.630(0.015) & 15.988(0.015) & 15.410(0.021) & 15.151(0.015) \\
$07^{\mathrm{h}}37^{\mathrm{m}}09\fs5$ & $-69\degr26\arcmin08\farcs7$ & 15.328(0.015) & 15.283(0.015) & 14.671(0.015) & 14.944(0.015) & 14.527(0.015) & 14.310(0.018) \\
$07^{\mathrm{h}}37^{\mathrm{m}}15\fs3$ & $-69\degr26\arcmin47\farcs6$ & 16.037(0.015) & 15.940(0.018) & 15.285(0.015) & 15.582(0.015) & 15.122(0.015) & 14.901(0.022) \\
$07^{\mathrm{h}}37^{\mathrm{m}}21\fs6$ & $-69\degr26\arcmin51\farcs1$ & 15.735(0.015) & 15.556(0.015) & 14.822(0.015) & 15.149(0.015) & 14.632(0.015) & 14.390(0.021) \\
$07^{\mathrm{h}}37^{\mathrm{m}}24\fs8$ & $-69\degr30\arcmin31\farcs8$ & 15.893(0.015) & 15.755(0.015) & 15.028(0.015) & 15.363(0.015) & 14.838(0.027) & 14.571(0.019) \\
$07^{\mathrm{h}}37^{\mathrm{m}}26\fs8$ & $-69\degr27\arcmin13\farcs2$ & 15.614(0.015) & 15.280(0.015) & 14.451(0.015) & 14.820(0.015) & 14.228(0.016) & 13.965(0.021) \\
$07^{\mathrm{h}}37^{\mathrm{m}}27\fs5$ & $-69\degr33\arcmin13\farcs9$ & --            & 17.418(0.021) & 16.330(0.015) & 16.829(0.018) & 15.947(0.018) & 15.572(0.016) \\
$07^{\mathrm{h}}37^{\mathrm{m}}29\fs8$ & $-69\degr33\arcmin01\farcs1$ & 18.177(0.015) & 17.834(0.015) & 16.905(0.015) & 17.318(0.015) & 16.632(0.018) & 16.316(0.017) \\
  \hline
 \end{tabular}
 \begin{tablenotes}
        \item Numbers in parenthesis correspond to 1\,$\sigma$ statistical uncertainties.
 \end{tablenotes}
\end{table*}

\begin{table*}
\caption{Optical photometry of SN\,2015F}
\label{op_phot_tab}
\begin{tabular}{lcccccccc}
   \hline
   \multicolumn{1}{l}{Date UT} &
   \multicolumn{1}{c}{MJD} &
   \multicolumn{1}{c}{$U$} &
   \multicolumn{1}{c}{$B$}  &
   \multicolumn{1}{c}{$V$} &
   \multicolumn{1}{c}{$g$} &
   \multicolumn{1}{c}{$r$} &
   \multicolumn{1}{c}{$i$} &
   \multicolumn{1}{c}{Tel.} \\
   \hline
2015-03-08 &  57089.07 & --            & --            & \textgreater19.017 & --       & --            & --            & 8 \\
2015-03-08 &  57089.18 & --            & --            & \textgreater18.709 & --       & --            & --            & 8 \\
2015-03-09 &  57090.12 & --            & --            & 18.055(0.101) & --            & --            & --            & 8 \\
2015-03-10 &  57091.80 & --            & 17.156(0.027) & 16.520(0.010) & 16.850(0.010) & 16.419(0.010) & 16.593(0.010) & 2 \\
2015-03-11 &  57092.00 & --            & --            & 16.456(0.010) & --            & --            & --            & 7 \\
2015-03-11 &  57092.99 & --            & --            & 15.857(0.019) & --            & --            & --            & 7 \\
2015-03-12 &  57093.81 & --            & 15.818(0.010) & 15.436(0.010) & 15.646(0.010) & 15.382(0.013) & 15.469(0.021) & 2 \\
2015-03-12 &  57093.90 & --            & 15.757(0.010) & 15.403(0.010) & 15.574(0.010) & 15.330(0.010) & --            & 1 \\
2015-03-13 &  57094.10 & --            & 15.682(0.010) & 15.312(0.010) & 15.494(0.010) & 15.250(0.010) & 15.315(0.010) & 6 \\
2015-03-13 &  57094.18 & --            & --            & 15.310(0.019) & --            & --            & --            & 7 \\
2015-03-13 &  57094.91 & 15.056(0.048) & --            & --            & --            & --            & --            & 1 \\
2015-03-14 &  57095.18 & --            & --            & 14.929(0.021) & --            & --            & --            & 7 \\
2015-03-14 &  57095.18 & 14.971(0.011) & 15.184(0.010) & 14.898(0.019) & 15.048(0.010) & 14.832(0.018) & --            & 6 \\
2015-03-14 &  57095.52 & 14.816(0.019) & 15.049(0.030) & 14.801(0.010) & 14.938(0.010) & 14.713(0.010) & 14.784(0.010) & 4 \\
2015-03-14 &  57095.78 & 14.708(0.013) & 14.915(0.010) & 14.702(0.010) & 14.805(0.010) & 14.620(0.010) & 14.696(0.024) & 1 \\
2015-03-14 &  57095.81 & 14.714(0.010) & 14.912(0.010) & 14.687(0.010) & 14.818(0.010) & 14.626(0.010) & 14.682(0.010) & 2 \\
2015-03-15 &  57096.85 & 14.336(0.010) & 14.562(0.010) & 14.387(0.014) & 14.496(0.011) & 14.315(0.010) & 14.391(0.010) & 2 \\
2015-03-16 &  57097.95 & 13.968(0.014) & 14.279(0.012) & 14.134(0.023) & 14.206(0.014) & 14.055(0.012) & 14.127(0.025) & 1 \\
2015-03-17 &  57098.79 & 13.865(0.010) & 14.085(0.014) & --            & 14.037(0.010) & 13.881(0.010) & 13.966(0.031) & 1 \\
2015-03-17 &  57098.82 & 13.876(0.063) & 14.077(0.010) & 13.945(0.010) & 14.040(0.010) & 13.879(0.010) & 13.979(0.010) & 2 \\
2015-03-18 &  57099.48 & 13.735(0.010) & 14.022(0.011) & 13.858(0.010) & 13.960(0.010) & 13.780(0.010) & 13.865(0.010) & 5 \\
2015-03-19 &  57100.84 & 13.504(0.010) & 13.749(0.010) & 13.640(0.010) & 13.706(0.010) & 13.584(0.010) & 13.711(0.010) & 1 \\
2015-03-20 &  57101.52 & 13.470(0.010) & 13.757(0.010) & 13.608(0.020) & 13.702(0.010) & 13.542(0.010) & 13.672(0.019) & 5 \\
2015-03-21 &  57102.53 & 13.369(0.014) & --            & --            & --            & --            & --            & 4 \\
2015-03-21 &  57102.77 & 13.368(0.014) & 13.600(0.010) & 13.461(0.010) & 13.536(0.010) & 13.426(0.010) & 13.628(0.010) & 3 \\
2015-03-22 &  57103.07 & 13.329(0.010) & 13.581(0.010) & 13.433(0.010) & 13.516(0.010) & 13.407(0.010) & 13.591(0.010) & 6 \\
2015-03-23 &  57104.85 & 13.275(0.010) & 13.461(0.010) & 13.335(0.010) & 13.401(0.010) & 13.317(0.010) & 13.619(0.027) & 1 \\
2015-03-23 &  57104.87 & 13.241(0.073) & --            & --            & --            & --            & --            & 3 \\
2015-03-24 &  57105.78 & 13.264(0.010) & 13.444(0.010) & 13.288(0.010) & 13.389(0.010) & 13.288(0.010) & 13.638(0.010) & 2 \\
2015-03-24 &  57105.85 & 13.255(0.026) & 13.457(0.010) & 13.293(0.010) & 13.373(0.010) & 13.283(0.012) & 13.625(0.012) & 1 \\
2015-03-25 &  57106.87 & 13.260(0.010) & 13.438(0.010) & 13.264(0.010) & 13.347(0.010) & 13.268(0.010) & 13.673(0.023) & 1 \\
2015-03-25 &  57106.90 & 13.306(0.024) & 13.486(0.033) & --            & --            & --            & --            & 3 \\
2015-03-26 &  57107.46 & 13.259(0.010) & 13.493(0.010) & 13.277(0.010) & 13.381(0.010) & 13.265(0.010) & 13.691(0.010) & 4 \\
2015-03-27 &  57108.40 & 13.313(0.010) & 13.507(0.011) & 13.281(0.010) & 13.385(0.010) & 13.285(0.010) & 13.737(0.010) & 4 \\
2015-03-27 &  57108.52 & --            & 13.551(0.013) & 13.300(0.010) & 13.438(0.010) & 13.297(0.010) & 13.740(0.010) & 5 \\
2015-03-28 &  57109.74 & --            & 13.533(0.010) & 13.268(0.010) & 13.399(0.010) & 13.263(0.010) & 13.745(0.030) & 1 \\
2015-03-28 &  57109.77 & --            & 13.569(0.010) & 13.295(0.010) & 13.438(0.010) & 13.292(0.010) & 13.776(0.010) & 2 \\
2015-03-29 &  57110.78 & --            & 13.604(0.010) & 13.291(0.010) & 13.457(0.010) & 13.287(0.010) & 13.782(0.010) & 2 \\
2015-03-30 &  57111.80 & --            & 13.710(0.010) & 13.321(0.010) & 13.501(0.010) & 13.325(0.011) & 13.818(0.010) & 3 \\
2015-03-30 &  57111.81 & 13.588(0.031) & 13.675(0.010) & 13.307(0.033) & 13.499(0.010) & 13.323(0.033) & 13.827(0.022) & 2 \\
2015-04-03 &  57115.74 & 14.016(0.049) & 14.081(0.010) & 13.502(0.010) & 13.767(0.010) & 13.585(0.021) & 14.070(0.010) & 2 \\
2015-04-07 &  57119.46 & 14.616(0.010) & --            & 13.766(0.012) & 14.092(0.010) & 13.817(0.010) & 14.197(0.015) & 5 \\
2015-04-08 &  57120.75 & --            & 14.712(0.010) & 13.825(0.010) & 14.202(0.010) & 13.843(0.010) & 14.181(0.023) & 1 \\
2015-04-08 &  57120.80 & --            & 14.738(0.016) & 13.884(0.021) & 14.321(0.214) & 13.844(0.010) & 14.159(0.028) & 3 \\
2015-04-08 &  57120.87 & --            & 14.724(0.010) & 13.835(0.010) & 14.219(0.010) & 13.857(0.010) & 14.166(0.026) & 2 \\
2015-04-08 &  57120.98 & --            & 14.778(0.010) & 13.875(0.011) & 14.258(0.014) & 13.876(0.010) & 14.163(0.010) & 6 \\
2015-04-10 &  57122.37 & --            & --            & 13.955(0.086) & 14.280(0.082) & --            & --            & 4 \\
2015-04-11 &  57123.02 & --            & 15.020(0.016) & 13.994(0.010) & 14.461(0.010) & 13.931(0.010) & 14.134(0.010) & 6 \\
2015-04-11 &  57123.38 & 15.047(0.047) & 15.067(0.010) & 14.010(0.010) & 14.508(0.010) & 13.947(0.010) & 14.107(0.010) & 5 \\
2015-04-12 &  57124.42 & --            & 15.164(0.012) & 14.094(0.010) & 14.611(0.010) & 13.982(0.010) & 14.125(0.018) & 4 \\
2015-04-13 &  57125.43 & --            & 15.301(0.036) & 14.156(0.010) & 14.716(0.014) & 14.023(0.010) & 14.100(0.010) & 5 \\
2015-04-14 &  57126.74 & --            & 15.418(0.010) & 14.209(0.010) & 14.827(0.010) & 14.021(0.010) & 14.118(0.021) & 1 \\
2015-04-14 &  57126.78 & --            & 15.456(0.010) & 14.239(0.010) & 14.855(0.010) & 14.052(0.010) & 14.130(0.010) & 2 \\
2015-04-16 &  57128.38 & --            & 15.594(0.010) & 14.331(0.022) & 14.967(0.027) & 14.101(0.010) & 14.110(0.013) & 5 \\
2015-04-16 &  57128.77 & 15.881(0.028) & 15.654(0.010) & 14.359(0.015) & 15.026(0.010) & 14.131(0.010) & 14.162(0.010) & 2 \\
2015-04-17 &  57129.79 & --            & 15.723(0.036) & 14.410(0.021) & 15.121(0.020) & 14.163(0.020) & 14.178(0.026) & 2 \\
2015-04-18 &  57130.80 & --            & 15.813(0.032) & 14.488(0.030) & 15.202(0.019) & 14.212(0.023) & 14.183(0.024) & 2 \\
2015-04-19 &  57131.82 & --            & 15.869(0.052) & 14.518(0.038) & --            & --            & --            & 3 \\
2015-04-19 &  57131.86 & --            & 15.859(0.070) & 14.538(0.039) & 15.302(0.023) & 14.265(0.016) & 14.205(0.051) & 2 \\
2015-04-20 &  57132.01 & --            & 15.904(0.027) & 14.547(0.022) & 15.298(0.019) & 14.274(0.031) & 14.175(0.026) & 6 \\
2015-04-20 &  57132.79 & 16.155(0.024) & 15.997(0.010) & 14.625(0.032) & 15.483(0.161) & 14.313(0.024) & --            & 3 \\
2015-04-20 &  57132.84 & 16.131(0.080) & --            & 14.594(0.010) & 15.354(0.010) & 14.299(0.032) & --            & 2 \\
\end{tabular}
\end{table*}

\begin{table*}
\contcaption{Optical photometry of SN\,2015F}
\begin{tabular}{lcccccccc}
   \hline
   \multicolumn{1}{l}{Date UT} &
   \multicolumn{1}{c}{MJD} &
   \multicolumn{1}{c}{$U$} &
   \multicolumn{1}{c}{$B$}  &
   \multicolumn{1}{c}{$V$} &
   \multicolumn{1}{c}{$g$} &
   \multicolumn{1}{c}{$r$} &
   \multicolumn{1}{c}{$i$} &
   \multicolumn{1}{c}{Tel.} \\
   \hline
2015-04-21 &  57133.05 & 16.120(0.045) & 15.982(0.025) & 14.622(0.010) & 15.396(0.010) & 14.334(0.010) & 14.223(0.010) & 6 \\
2015-04-21 &  57133.76 & --            & 16.000(0.026) & 14.650(0.048) & 15.444(0.024) & 14.380(0.023) & --            & 2 \\
2015-04-21 &  57133.79 & --            & 16.004(0.010) & 14.669(0.010) & 15.453(0.010) & 14.366(0.010) & --            & 3 \\
2015-04-21 &  57133.97 & --            & 16.027(0.033) & 14.687(0.022) & 15.459(0.026) & 14.386(0.023) & 14.267(0.030) & 6 \\
2015-04-22 &  57134.71 & --            & 16.147(0.032) & 14.764(0.024) & 15.505(0.023) & 14.452(0.027) & 14.300(0.021) & 3 \\
2015-04-24 &  57136.11 & 16.312(0.036) & 16.178(0.010) & 14.854(0.015) & 15.616(0.010) & 14.549(0.010) & 14.402(0.023) & 6 \\
2015-04-27 &  57139.42 & 16.398(0.010) & 16.354(0.010) & 15.057(0.010) & 15.800(0.010) & 14.824(0.010) & 14.694(0.010) & 4 \\
2015-05-01 &  57143.71 & --            & --            & 15.315(0.055) & 16.045(0.055) & --            & --            & 2 \\
2015-05-02 &  57144.81 & 16.668(0.094) & 16.625(0.043) & 15.319(0.020) & 16.025(0.010) & 15.111(0.010) & 15.011(0.010) & 3 \\
2015-05-05 &  57147.41 & --            & 16.519(0.047) & 15.350(0.025) & 16.049(0.027) & 15.191(0.022) & 15.091(0.027) & 4 \\
2015-05-08 &  57150.74 & 16.773(0.033) & 16.644(0.021) & 15.456(0.012) & 16.130(0.010) & 15.285(0.010) & 15.229(0.026) & 1 \\
2015-05-09 &  57151.77 & --            & 16.811(0.059) & 15.508(0.022) & --            & --            & --            & 3 \\
2015-05-09 &  57151.81 & 16.686(0.072) & 16.695(0.023) & 15.504(0.010) & 16.164(0.010) & 15.342(0.010) & 15.281(0.013) & 1 \\
2015-05-10 &  57152.06 & 16.853(0.055) & 16.722(0.017) & 15.536(0.013) & 16.209(0.010) & 15.382(0.010) & 15.321(0.020) & 6 \\
2015-05-10 &  57152.73 & --            & 16.722(0.035) & 15.530(0.019) & 16.187(0.021) & 15.399(0.021) & 15.319(0.022) & 2 \\
2015-05-10 &  57152.79 & --            & 16.745(0.020) & 15.547(0.010) & 16.189(0.010) & 15.392(0.010) & --            & 3 \\
2015-05-10 &  57152.96 & --            & 16.765(0.036) & 15.563(0.018) & 16.211(0.020) & 15.399(0.023) & 15.337(0.022) & 6 \\
2015-05-14 &  57156.35 & --            & 16.795(0.053) & 15.614(0.029) & 16.220(0.032) & 15.463(0.026) & 15.431(0.034) & 5 \\
2015-05-16 &  57158.74 & --            & 16.829(0.010) & 15.648(0.010) & 16.293(0.010) & 15.560(0.030) & --            & 3 \\
2015-05-17 &  57159.37 & 16.945(0.019) & 16.800(0.010) & 15.690(0.010) & 16.281(0.010) & 15.583(0.011) & 15.561(0.023) & 5 \\
2015-05-18 &  57160.34 & --            & 16.823(0.051) & 15.717(0.021) & 16.276(0.025) & 15.601(0.026) & 15.579(0.030) & 4 \\
2015-05-22 &  57164.98 & --            & 16.924(0.056) & 15.837(0.031) & 16.405(0.042) & 15.811(0.050) & --            & 6 \\
2015-05-23 &  57165.40 & 16.986(0.090) & 16.923(0.023) & 15.869(0.024) & 16.391(0.015) & 15.781(0.010) & 15.781(0.026) & 4 \\
2015-05-27 &  57169.69 & --            & 17.024(0.068) & 15.991(0.028) & --            & --            & --            & 3 \\
2015-05-27 &  57169.73 & --            & 16.966(0.035) & 15.962(0.025) & 16.478(0.022) & 15.943(0.025) & 15.972(0.026) & 2 \\
2015-05-28 &  57170.76 & --            & 16.919(0.010) & 15.910(0.023) & 16.495(0.026) & 15.897(0.010) & --            & 1 \\
2015-06-04 &  57177.34 & --            & 17.072(0.055) & 16.126(0.025) & 16.596(0.027) & 16.169(0.031) & 16.209(0.030) & 4 \\
2015-06-05 &  57178.96 & --            & 17.134(0.049) & 16.214(0.022) & 16.645(0.017) & 16.242(0.015) & 16.263(0.029) & 6 \\
2015-06-07 &  57180.95 & --            & 17.153(0.049) & 16.246(0.030) & 16.690(0.025) & 16.267(0.029) & 16.371(0.041) & 6 \\
2015-06-08 &  57181.72 & --            & 17.166(0.031) & 16.276(0.010) & 16.664(0.013) & 16.311(0.026) & --            & 3 \\
2015-06-09 &  57182.72 & --            & 17.159(0.031) & 16.255(0.026) & 16.639(0.017) & 16.328(0.014) & --            & 1 \\
2015-06-10 &  57183.37 & --            & 17.077(0.075) & 16.248(0.016) & 16.736(0.227) & 16.369(0.030) & 16.345(0.010) & 5 \\
2015-06-11 &  57184.95 & --            & 17.145(0.054) & 16.346(0.029) & 16.717(0.024) & 16.402(0.027) & 16.469(0.040) & 6 \\
2015-06-17 &  57190.70 & --            & --            & --            & 16.784(0.010) & 16.598(0.058) & --            & 3 \\
\hline
\end{tabular}
\begin{tablenotes}
\item Numbers in parenthesis correspond to 1\,$\sigma$ statistical uncertainties.
\item  Telescopes: 1=1m\,LCOGT--10/SAAO; 2=1m\,LCOGT--12/SAAO; 3=1m\,LCOGT--13/SAAO; 4=1m\,LCOGT--03/SSO; 5=1m\,LCOGT--11/SSO; 6=1m\,LCOGT--05/CTIO; 7=EFOSC--NTT/La Silla; 8=PROMPTs/CTIO.
\end{tablenotes}
\end{table*}

\begin{center}
\begin{table*}
\caption{$JHK_{s}$ photometric sequence around SN\,2015F.}
\label{nir_local_seq_tab}
\begin{tabular}{@{}ccccc}
\hline
R.A. & Decl. & $J$ & $H$ & $K_{s}$ \\
\hline
$07^{\mathrm{h}}35^{\mathrm{m}}47\fs2$ & $-69\degr30\arcmin55\farcs2$ & $14.633$($0.006$) & $14.264$($0.019$) & $14.123$($0.034$) \\
$07^{\mathrm{h}}36^{\mathrm{m}}31\fs2$ & $-69\degr29\arcmin59\farcs1$ & $14.866$($0.032$) & $14.547$($0.030$) & $14.366$($0.048$) \\
$07^{\mathrm{h}}36^{\mathrm{m}}39\fs1$ & $-69\degr30\arcmin44\farcs3$ & $13.358$($0.017$) & $12.796$($0.017$) & $12.629$($0.019$) \\
  \hline
 \end{tabular}
 \begin{tablenotes}
        \item Numbers in parenthesis correspond to 1\,$\sigma$ statistical uncertainties.
 \end{tablenotes}
\end{table*}
\end{center}

\begin{table*}
\caption{$JHK_{s}$ Photometry of SN\,2015F.}
\label{nir_phot_tab}
\begin{tabular}{@{}ccccc}
\hline
Date UT & MJD & $J$ & $H$ & $K_{s}$ \\
\hline
2015-03-12 & $57093.17$ & $15.092$($0.014$) & $14.981$($0.036$) & $14.939$($0.046$) \\
2015-03-12 & $57094.00$ & $14.771$($0.017$) & $14.668$($0.032$) & $14.601$($0.065$) \\
2015-03-14 & $57095.03$ & $14.441$($0.021$) & $14.431$($0.024$) & $14.222$($0.027$) \\
2015-03-19 & $57100.03$ & $13.409$($0.023$) & $13.466$($0.023$) & $13.364$($0.027$) \\
2015-03-21 & $57102.04$ & $13.329$($0.033$) & $13.637$($0.028$) & $13.271$($0.035$) \\
2015-03-29 & $57110.12$ & $13.631$($0.013$) & $13.703$($0.012$) & $13.416$($0.057$) \\
2015-04-10 & $57122.12$ & $14.652$($0.010$) & $13.519$($0.010$) & $13.481$($0.029$) \\
2015-04-12 & $57124.06$ & $14.643$($0.010$) & $13.508$($0.010$) & $13.512$($0.016$) \\
2015-04-17 & $57129.03$ & $14.399$($0.010$) & $13.449$($0.027$) & $13.404$($0.010$) \\
2015-04-27 & $57139.98$ & $14.598$($0.010$) & $13.941$($0.010$) & $14.037$($0.025$) \\
  \hline
 \end{tabular}
 \begin{tablenotes}
        \item Numbers in parenthesis correspond to 1\,$\sigma$ statistical uncertainties.
 \end{tablenotes}
\end{table*}

\bsp	
\label{lastpage}
\end{document}